\documentclass[%
 reprint,longbibliography,preprintnumbers,
nofootinbib,
 amsmath,amssymb,
 aps,
pre,
]{revtex4-1}
\UseRawInputEncoding
\pdfoutput=1
\usepackage{flushend}
\usepackage{dcolumn}
\usepackage{bm}
\usepackage{balance}
\usepackage{tensor}
\usepackage{physics}
\usepackage{wrapfig}
\usepackage{subcaption}
\usepackage{graphicx}
\usepackage{mathtools}
\usepackage[thinc]{esdiff}

\usepackage[normalem]{ulem}

\usepackage[colorlinks = true,
            linkcolor = blue,
            urlcolor  = blue,
            citecolor =green,
            anchorcolor = blue]{hyperref}
\usepackage{verbatim}
\usepackage{color,ulem}
\usepackage[english]{babel}

\usepackage[utf8]{inputenc}

\newcommand{\beq}{\begin{eqnarray}}
\newcommand{\eeq}{\end{eqnarray}}

\usepackage{amsmath}
\usepackage{tikz}
\usetikzlibrary{decorations.pathmorphing}
\usetikzlibrary{shapes.misc}
\tikzset{cross/.style={cross out, draw=black, minimum size=8*(#1-\pgflinewidth), inner sep=0pt, outer sep=0pt},
cross/.default={1pt}}
\usetikzlibrary{patterns,math}
\begin{document}

\title{Topological Bardeen–Cooper–Schrieffer theory of superconducting quantum rings}

\author{Elena Landrò$^{1,4}$}%
\email{emglandro@studenti.uninsubria.it}
\author{Vladimir M. Fomin$^{2,3}$}%
\email{v.fomin@ifw-dresden.de}
\author{Alessio Zaccone$^{1}$}%
\email{alessio.zaccone@unimi.it}

\vspace{1cm}

\affiliation{$^{1}$Department of Physics ``A. Pontremoli'', University of Milan, via Celoria 16, 20133 Milan, Italy}
\affiliation{$^2$ 
Institute for Emerging Electronic Technologies
Leibniz IFW Dresden
Helmholtzstrasse 20, D-01069 Dresden, Germany}
\affiliation{$^3$ 
Faculty of Physics and Engineering,
Moldova State University,
strada Alexei Mateevici 60, MD-2009 Chişin\u{a}u, Republic of Moldova}
\affiliation{$^{4}$Department of Science and High Technology, Insubria University, Via Valleggio 11, Como, 22100 Como, Italy}

\begin{abstract}
Quantum rings have emerged as a playground for quantum mechanics and topological physics, with promising technological applications. Experimentally realizable quantum rings, albeit at the scale of a few nanometers, are 3D nanostructures. Surprisingly, no theories exist for the topology of the Fermi sea of quantum rings, and a microscopic theory of superconductivity in nanorings is also missing. In this paper, we remedy this situation by developing a mathematical model for the topology of the Fermi sea and Fermi surface, which features non-trivial hole pockets of electronic states forbidden by quantum confinement, as a function of the geometric parameters of the nanoring. The exactly solvable mathematical model features two topological transitions in the Fermi surface upon shrinking the nanoring size either, first, vertically (along its axis of revolution) and, then, in the plane orthogonal to it, or the other way round. These two topological transitions are reflected in a kink and in a characteristic discontinuity, respectively, in the electronic density of states (DOS) of the quantum ring, which is also computed. Also, closed-form expressions for the Fermi energy as a function of the geometric parameters of the ring are provided. These, along with the DOS, are then used to derive BCS equations for the superconducting critical temperature of nanorings as a function of the geometric parameters of the ring. The $T_c$ varies non-monotonically with the dominant confinement size and exhibits a prominent maximum, whereas it is a monotonically increasing function of the other, non-dominant, length scale. For the special case of a perfect square toroid (where the two length-scales coincide), the $T_c$ increases monotonically with increasing the confinement size, and in this case, there is just one topological transition.
\end{abstract}

\maketitle
\section{Introduction}
Quantum rings have emerged as a playground for both fundamental quantum mechanics and technological applications \cite{Fomin_book,Fomin_book2}. A wealth of fundamental phenomena can be hosted in nanometric conducting rings, ranging from quantum interference effects (e.g., the Aharonov-Bohm effect)\cite{PhysRevLett.99.146808} and magnetoresistance oscillations \cite{PhysRevB.30.4048} to superconductivity \cite{PhRX_2024}. Experimentally, quantum rings of various materials have been fabricated by diverse methods, including lithography \cite{Levy1990,Chandrasekhar1991,Mailly1993}, self-assembly through partial overgrowth \cite{Garcia1997}, droplet etching \cite{Sanguinetti2011,Wu2011, Heyn2011}, focused electron-beam-based 3D printing \cite{Plank2019} or thermochemical edge reconfiguration from a nanoplatelet \cite{Salzmann_2021}. 

Much theoretical research on quantum rings has been devoted to studying quantum interference effects \cite{Fomin2007,PhysRevB.78.245315} as well as optical (e.g., excitonic) effects \cite{Fomin2008,GRBIC20081273} and transport properties \cite{Papari2022}.
For example, nanorings, also referred to as quantum doughnuts, have been shown to be able to slow down, and even freeze, the propagation of light \cite{PhysRevLett.102.096405}.

In spite of this intense research, especially in the context of analytically solvable 1D and quasi-1D models and numerical analysis of realistic quantum rings of complex geometry and materials composition, some basic properties of quantum rings are still lacking a systematic theoretical description. For example, the mathematical description of the Fermi surface and of the Fermi sea topology of metallic nanorings, as well as a microscopic theory of superconductivity under 3D nanoring confinement, are currently missing.
We remedy this situation in the present paper by presenting a mathematical theory of the free electron gas in quantum rings, with application to 3D metallic nanorings, and a microscopic theory of superconductivity thereof. 
Building upon a recent quantum confinement model for nanometric thin films \cite{Travaglino2023}, we derive analytical closed-form expressions for the Fermi energy as a function of the geometric parameters of the quantum rings, and we also derive analytical expressions for the corresponding density of states (DOS), for the Fermi energy and for the superconducting critical temperature. 

Two topological transitions driven by the 3D ring confinement are predicted, upon varying the geometric parameters of the nanoring. The first transition occurs when one of the two confinement lengths of the ring, i.e. either the vertical or the in-plane one, shrinks to such an extent that the corresponding hole pockets cross the Fermi level.
This transition marks the change from the spherical Fermi surface of the bulk material to a non-trivial Fermi surface with fundamental homotopy group $\mathbb{Z}$. The second transition occurs when also the second confinement length scale crosses the Fermi level.
This transition marks the change from the $\mathbb{Z}$ Fermi surface to another topologically non-trivial Fermi surface with fundamental homotopy group $\mathbb{Z}_6$. 
In the special case of a perfect square toroid, instead, we have only one confining length scale. In this case, there is only one topological transition, from the trivial spherical surface to a non-trivial surface topology with group $\mathbb{Z}_6$.
We will also show how these topological transitions affect the shape of the electronic density of states and of the superconducting critical temperature as a function of the ring geometry.

These findings may have important implications for the physical properties of nanorings, ranging from superconductivity (as discussed below) to magnetoresistance. For the latter, one can recall that the relative magnetoresistance goes as $\sim (\omega_c \tau)^2$ \cite{Kittel}, where $\omega_c$ is the cyclotron frequency and $\tau$ is the collisional mean free time of the electrons in the system. $\tau$ depends also on the Fermi velocity $v_F$, which, in turn, depends on the Fermi energy $\epsilon_F$. Hence, it is clear that the possibility of tuning $\epsilon_F$ by the geometry of the nanoring according to the formulae derived in this paper may open up a new way to understand and tailor the geometry-dependent magnetoresistance of nanorings.

\section{Previous approaches: thin films}
\begin{figure}[h!]
    \centering
    \includegraphics[width=0.45\textwidth]{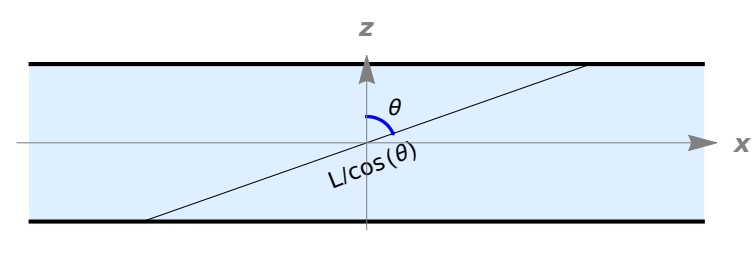}
    \caption{2D section of a thin film of thickness $L$, confined along $z$ direction and infinite along the $y$ and $x$ directions. A free electron (quantum plane wave) is assumed to have a maximum
wavelength equal to the length of the medium in the direction
of motion, which can be expressed as a function of the angle
$\theta$, thanks to the cylindrical symmetry,  as $\lambda_{max} = L/ \cos{\theta}$. This
leads to a cutoff in the accessible values of the wavevector $k$. Adapted with permission from Ref. \cite{Phillips_2021}.}
    \label{fig:travaglino1}
\end{figure}

When a quantum system is confined, its fundamental properties change due to the rearrangement of the accessible states in momentum space. While numerical models provide insights, they often overshadow physical mechanisms and rely on artificial boundary conditions, which are scarcely relevant for experimental comparison \cite{Blatt,Valentinis}. Here we use a theoretical approach that applies to nanosystems with spatial dimensions greater than two. This approach, for non-interacting quasi-particles, was introduced in Ref. \cite{Travaglino_2022} for bosons and in Ref. \cite{Travaglino2023} for free electrons, and in both cases, a thin film geometry was considered, as schematically depicted in Fig. (\ref{fig:travaglino1}).

\begin{figure}
    \centering
    \includegraphics[width=0.45\textwidth]{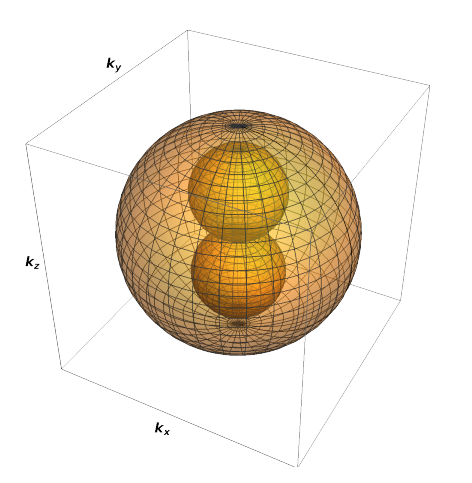}
    \caption{The allowed momentum space for phonon plane waves propagating in a confined sample with the thin film geometry sketched in Fig. (\ref{fig:travaglino1}).
The two inner spheres represent the set of forbidden states in k-space (hole pockets), while the outer sphere is the Fermi sphere for the bulk material. The volume of available states in k-space is represented by the volume of the outer Fermi sphere minus the volumes of the two smaller spheres which represent states that are not available due to confinement. Adapted with permission from Ref. \cite{Phillips_2021}.}
    \label{fig:travaglino2}
\end{figure}

Confinement effects are incorporated by setting a thickness dependent cut-off on low-energy states, where the maximum wavelength of particles is limited by the sample's geometry. This condition dictates a minimum wavenumber for free quantum particles, facilitating analytical calculations of momentum space geometry and topology. In particular, in the thin film case-study, it was found, analytically, that the k-space geometry of quasiparticles has the topology shown schematically in Fig. (\ref{fig:travaglino2}).

The corresponding density of states (DOS) was also computed analytically for the free electron gas in a thin film, as a function of energy. The obtained forms of the DOS are as follows \cite{Travaglino2023}:
\begin{equation}
    g(\epsilon)= \begin{cases}
        \frac{Vg_sLm^2}{2\pi^3 \hbar^4}\epsilon \quad & \text{if  } \epsilon < \frac{2\pi^2 \hbar^2}{mL^2}\\
        \frac{V g_s (2m)^{3/2}}{(2\pi)^2\hbar^3}\epsilon^{1/2}\quad & \text{if  } \epsilon > \frac{2\pi^2 \hbar^2}{mL^2}
    \end{cases}
\end{equation}

The key outcome resulting from the intersection of the two density of states regimes mentioned above is the displacement of the Fermi level with the film thickness. This shift occurs because certain low-energy states become inaccessible as the film thickness is lowered. This shift gains significance when one applies the model to superconductors, where it directly controls the superconducting critical temperature $T_c$ \cite{Travaglino2023}. The Fermi energy was computed by imposing a total electron number $N$ in the sample of volume $V$ (the electron number density is denoted by $\rho=N/V$). The result was given by \cite{Travaglino2023}:
\begin{equation}
     \epsilon_F= \begin{cases}
        \frac{\hbar^2}{2m}\bigg(3\pi^2\rho\bigg)^{2/3}\bigg(1+ \frac{2}{3}\frac{\pi}{\rho L^3}\bigg)^{2/3} \quad & \text{if } L > \bigg(\frac{2\pi^2}{\rho}\bigg)^{1/3}\\
        \frac{\hbar^2}{m}\bigg[\frac{(2\pi)^3\rho}{L}\bigg]^{1/2}\quad & \text{if } L < \bigg(\frac{2\pi^2}{\rho}\bigg)^{1/3}.
    \end{cases}
\end{equation}

 This result supports the intuition that, for films that are not too thin, the Fermi level is shifted 
 upwards because of the significant number of lower-energy forbidden states due to confinement, hence forcing more states to be accommodated at the Fermi surface. This leads to higher values of the chemical potential.
Upon further increasing the confinement (decreasing the film thickness $L$), the two spheres describing the hole pockets of prohibited states (cfr. Fig. \ref{fig:travaglino2}) eventually intersect the Fermi sphere. This marks a topological transition in the Fermi surface, from the trivial sphere $\pi_{1}(S^{2})=0$ to a non-trivial Fermi surface with homotopy group 
$\pi_{1} \simeq \pi_{1}(S^{1})=\mathbb{Z}$. From this point onwards, the available surface for distributing fermions increases, resulting in a decrease in the density of states at the Fermi level, while the Fermi energy increases with decreasing the thickness $L$ with a weaker square root law.

\section{Confinement model for quantum rings and the topology of Fermi sphere}

\begin{figure}[h]
\centering
    \includegraphics[width=0.4\textwidth]{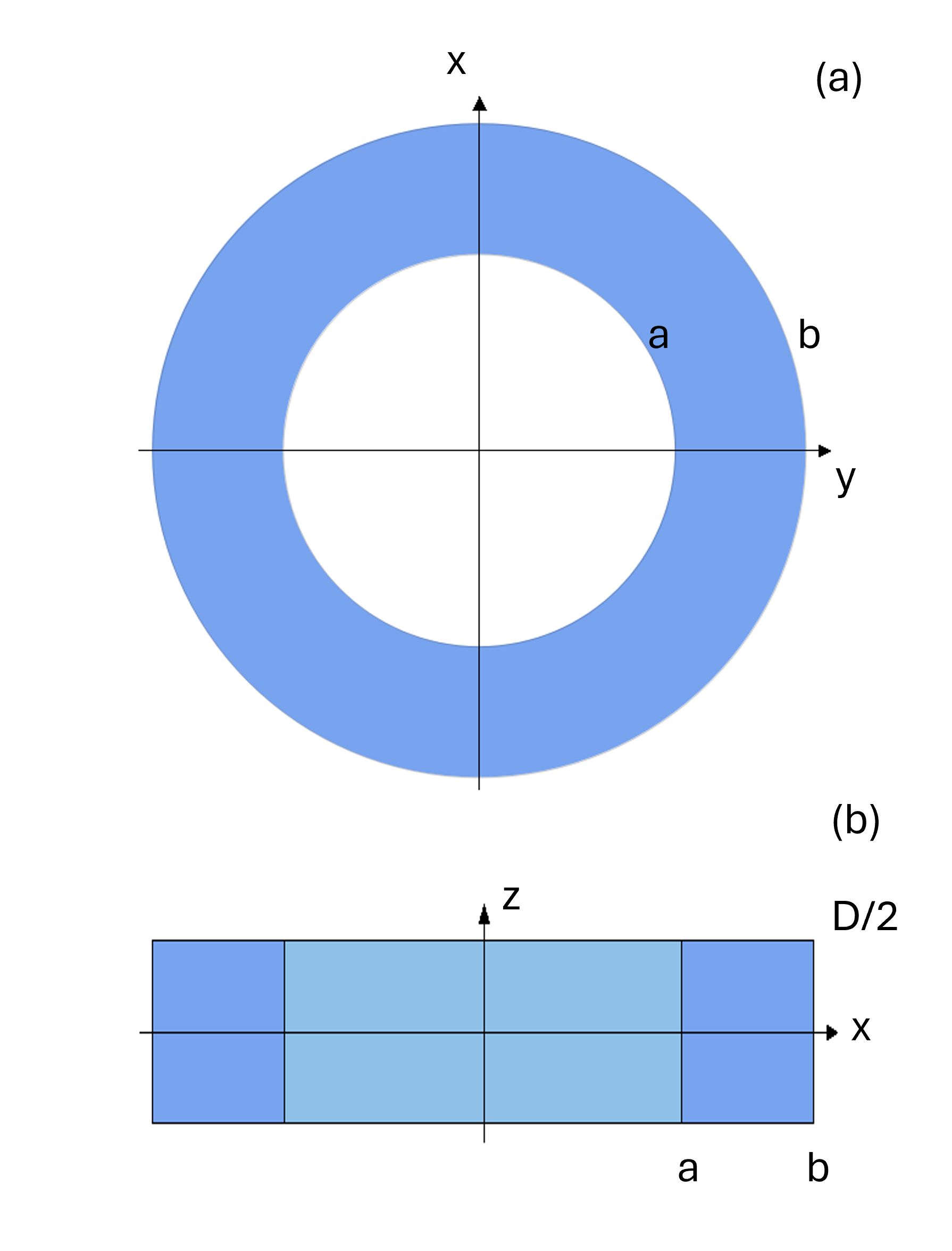}
\caption{Schematic section in real space of the confined ring sample. (a) Projection of the confinement in xy space; a ring of inner radius $a$ and outer radius $b$.  (b) Projection of the confinement in the xz plane: a rectangle with dimensions $b$ and $D$. The inner light blue rectangle with dimensions $a$ and $D$ depicts the area in the xy plane where the inner circle is located.}
\label{fig:real-space}
\end{figure}

Modern technologies, e.g., cross-sectional scanning tunneling microscopy, have allowed for atomic-scale characterization of quantum rings and unveiled their complicated confinement geometry \cite{Offermans2005}. We consider here a simplified model of a quantum ring as a hollow hard-wall cylinder shown in Fig. \ref{fig:real-space}. Its geometric parameters must be adjusted to experimental ones.

We have a ring-shaped confinement in the xy plane and a confinement along the z-direction.
The parameters of the system are therefore:
\begin{enumerate}
    \item the radii that define the ring confinement; Let us call $a$ and $b$ the radius of the inner and outer circle, respectively.
    \item the linear parameter that defines the confinement along $z$; let us call it $D$.
\end{enumerate}

In order to be able to provide analytical results, we treat the electron as a free particle with energy $\epsilon = \frac{\hbar^2k^2}{2m}$
as customary for good metals.

Confinement places limits on the maximum possible wavelength, and therefore on the minimum possible wave-number (because they are inversely related) and imposes a cut-off in the accessible low-energy states.

\begin{figure}
    \centering
    \includegraphics[width=0.4\textwidth]{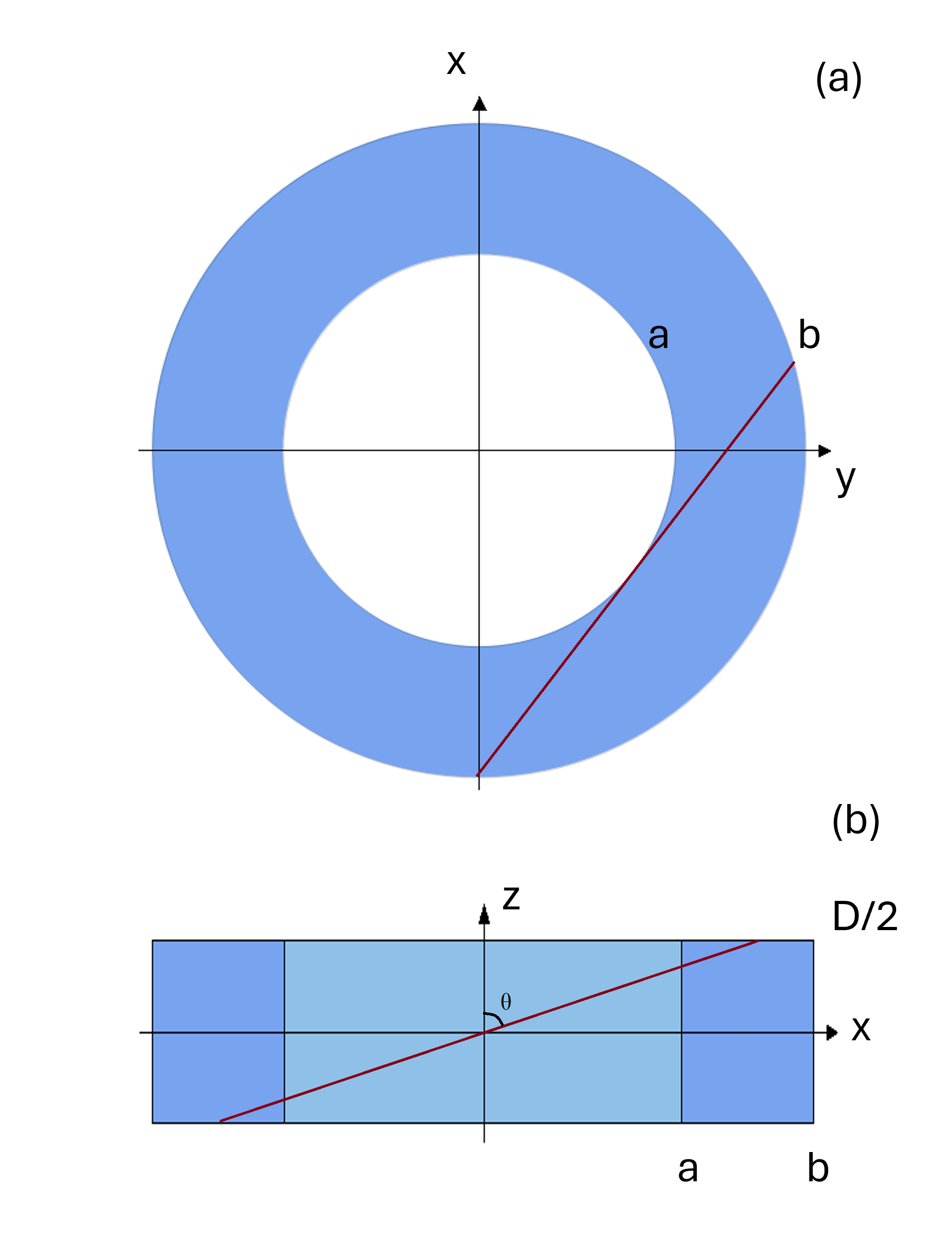}
    \caption{Schematic section in real space of the confined ring sample. (a) Projection of the confinement in xy space; the red line represents the path with maximum length. (b) Projection of the confinement in xz space; the red line represents the path with maximum length. $\theta$ is the angle between the red line and the z-axis.}
    \label{fig:realSpaceLine}
\end{figure}

To compute the maximum free-electron wavelength allowed by the confinement in the ring, we have to find the modulus of the line of maximum length contained entirely in the ring.
Thanks to the cylindrical geometry of the system, the problem can be broken down into two parts:\\
(i) Find the maximum length contained in the xy plane;\\
(ii) Compute the confinement also in the third dimension.\\
It is easy to see that the line with maximum modulus in the xy plane is the tangent to the inner circumference, as shown in
Fig. \ref{fig:realSpaceLine}.

The modulus $\mathcal{L}$ of that line is given by:
\begin{equation}
    \mathcal{L} = 2\sqrt{b^2-a^2}.
\end{equation}

To add the third dimension, we define the angle $\theta$ as shown in Fig. \ref{fig:realSpaceLine}.

By studying limiting cases in $\theta$, it is easy to see that we have confinement in the xy plane near $\theta = \frac{\pi}{2}$ and along z close to $\theta = 0$. We thus have the following relations:
\[  \lambda_{max} \equiv
\begin{cases}
\frac{\mathcal{L}}{\sin(\theta)}  
   & \text{if }   \big|\tan(\theta)\big| \geq \frac{\mathcal{L}}{D}\\
\frac{D}{\cos(\theta)}
   &\text{if }  \big|\tan(\theta)\big| \leq \frac{\mathcal{L}}{D}.
\end{cases}
\]

We notice that if $a=0$ we find the equations obtained in \cite{Phillips_2021}.

The constraints on the wave number are therefore:
\[ k_{min} \equiv
\begin{cases}
2\pi \frac{\sin(\theta)}{\mathcal{L}} 
   & \text{if }  \big|\tan(\theta)\big| \geq \frac{\mathcal{L}}{D}\\
2\pi \frac{\cos(\theta)}{D}
   &\text{if } \big|\tan(\theta)\big| \leq \frac{\mathcal{L}}{D}.
\end{cases}
\]

\begin{figure}[h]
    \centering
    \includegraphics[width=0.35\textwidth]{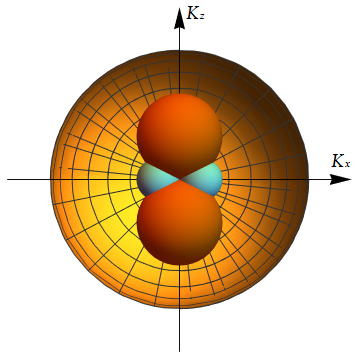}
    \caption{ Rendering of the geometry of the different regions of the k-space. The yellow sphere is the Fermi sphere, while the orange and the green spheres are the regions in which are located the hole pockets of forbidden states. This is not to scale.}
    \label{fig:rendering}
\end{figure}

It is now possible to analytically calculate the geometry of the corresponding volume in momentum space.\\
In Fig. \ref{fig:rendering} we can see, inside the Fermi sphere, the existence of 4 spheres that represent the forbidden states in momentum space that are inaccessible due to the confinement (hole pockets). The four spheres of hole pockets can be distinguished into a pair of spheres which are symmetric across the origin along the $k_z$ axis, and a pair of spheres symmetric along the $k_x$ axis. This therefore identifies two different values for the radii of the two distinct pairs.

The radii of these spheres (hole pockets) are functions of the geometrical parameters of the ring; as they change, the number of low energy states and, consequently, the density of states (DOS) will change, for the free electron case.
The DOS, indeed, can be expressed as:
\begin{equation}
    g(\epsilon)=  \diff{}{\epsilon}N(\epsilon'<\epsilon),\label{DOS}
\end{equation}
where $N(\epsilon'<\epsilon)$ is the number of states having energy lower than $\epsilon$ (this can be computed by knowing the volume of occupied states in k-space and applying the dispersion relation between $\epsilon$ and $k$).

We can identify three cases:

\hfill \break
\textbf{1) the Fermi radius is greater than both diameters that define the hole pockets of forbidden states.}

\begin{figure}[h]
    \centering
    \includegraphics[width=0.3\textwidth]{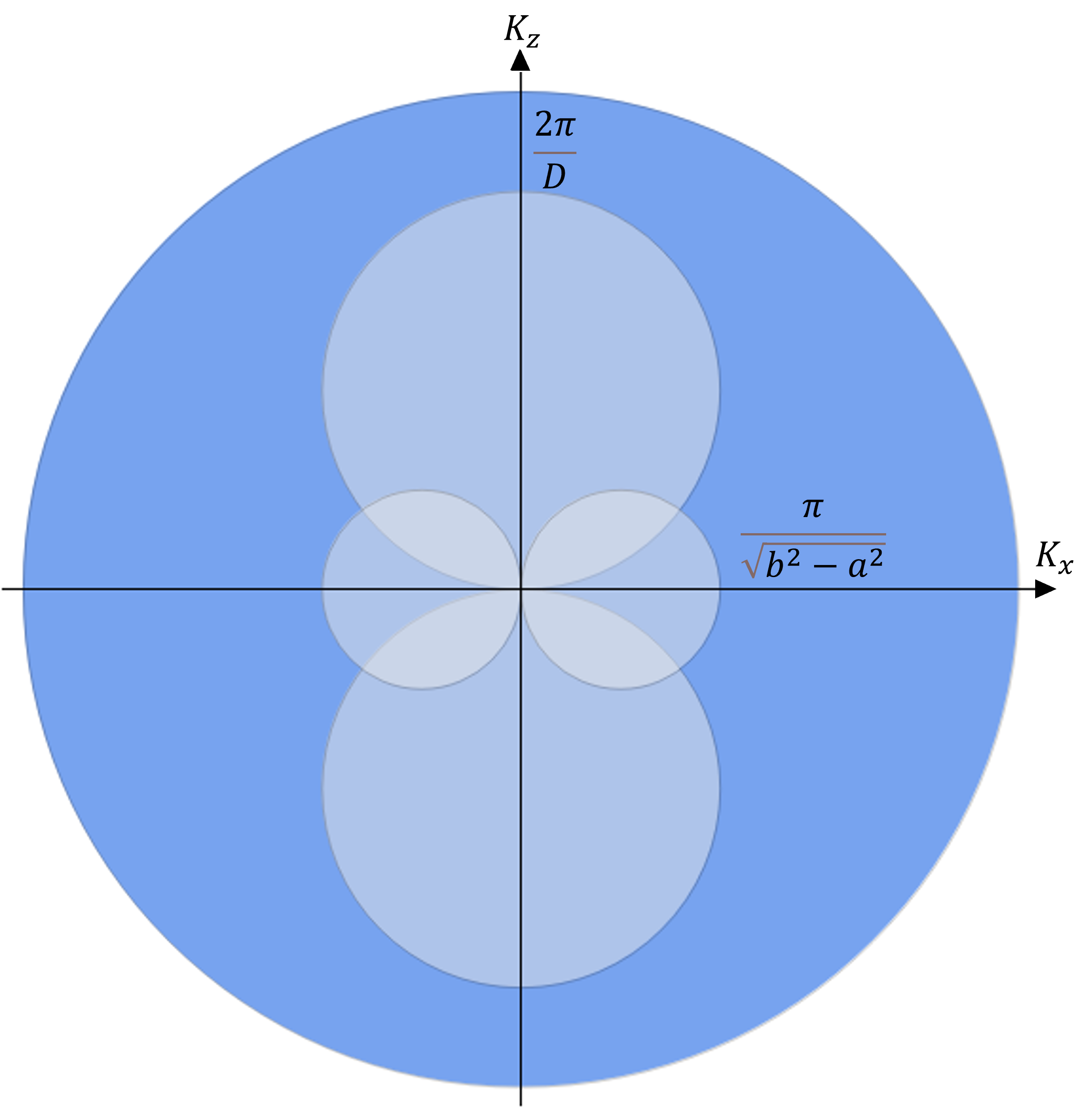}
    \caption{Projection of the 3D rendering of Fig. \ref{fig:rendering} in the $k_x-k_z$ space. In this case, the Fermi radius is greater than both diameters that define the hole pockets of forbidden states.}
    \label{fig:proj1}
\end{figure}

This case is defined by the following system of inequalities:

\[\begin{cases}
    k_{F} >\frac{2\pi}{\mathcal{L}}\\
    k_{F} >\frac{2\pi}{D}.
\end{cases}
\]
In Fig. \ref{fig:proj1} the corresponding situation is presented as a 2D projection on the $k_z$-$k_x$ plane.
\\

The fundamental homotopy group is therefore a point, or $\pi_1(S^2)=0$.

\hfill \break
\textbf{2) One of the diameters that define the hole pockets (forbidden states) is greater than the Fermi radius.}

This case is defined by the following system of inequalities:
\[
\begin{cases*}
  k_{F}< \frac{2\pi}{\mathcal{L}} \\
  k_{F}> \frac{2\pi}{D}
\end{cases*} \quad \text{or} \quad
\begin{cases*}
  k_{F}> \frac{2\pi}{\mathcal{L}} \\
  k_{F}< \frac{2\pi}{D}. 
\end{cases*}
\]

\begin{figure}[h!]
    \centering
    \includegraphics[width=0.3\textwidth]{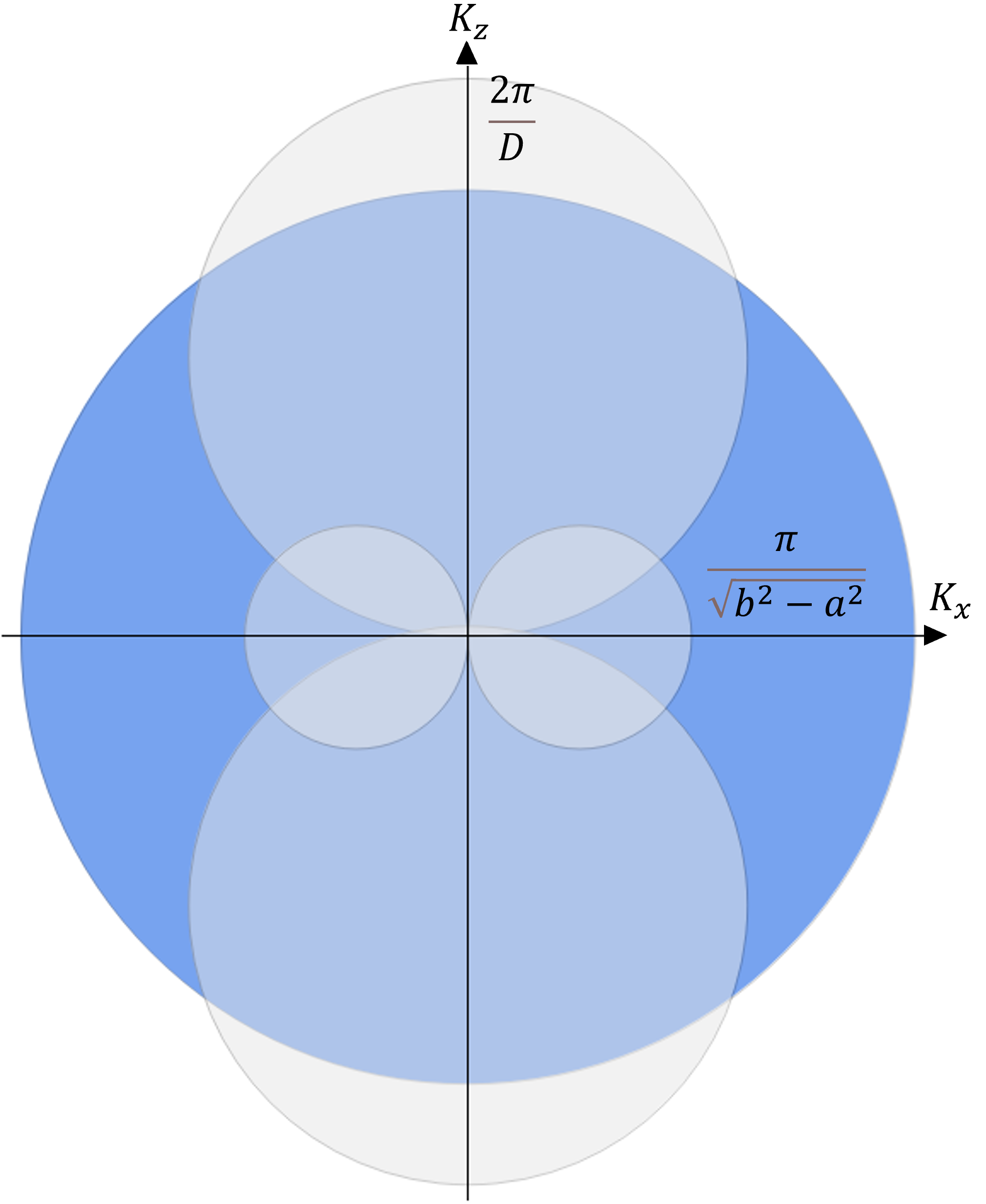}
    \caption{Projection of the 3D rendering of Fig. \ref{fig:rendering} in $k_x-k_z$ space. In this case, only one diameter that defines the hole pockets of forbidden states is greater than the Fermi radius.}
    \label{fig:proj2}
\end{figure}

In Fig. \ref{fig:proj2} the corresponding situation is presented as a 2D projection on the $k_z$-$k_x$ plane.
\\

The fundamental homotopy group, in this case, is $\mathbb{Z}$.

\hfill \break
\hfill

\textbf{3) Both diameters that define the hole pockets of forbidden states are greater than the Fermi radius. }

This case is defined by the following system of inequalities:
\[
\begin{cases}
  k_{F} <\frac{2\pi}{\mathcal{L}}\\
  k_{F} <\frac{2\pi}{D}.
\end{cases}
\]

\begin{figure}[h!]
    \centering
    \includegraphics[width=0.3\textwidth]{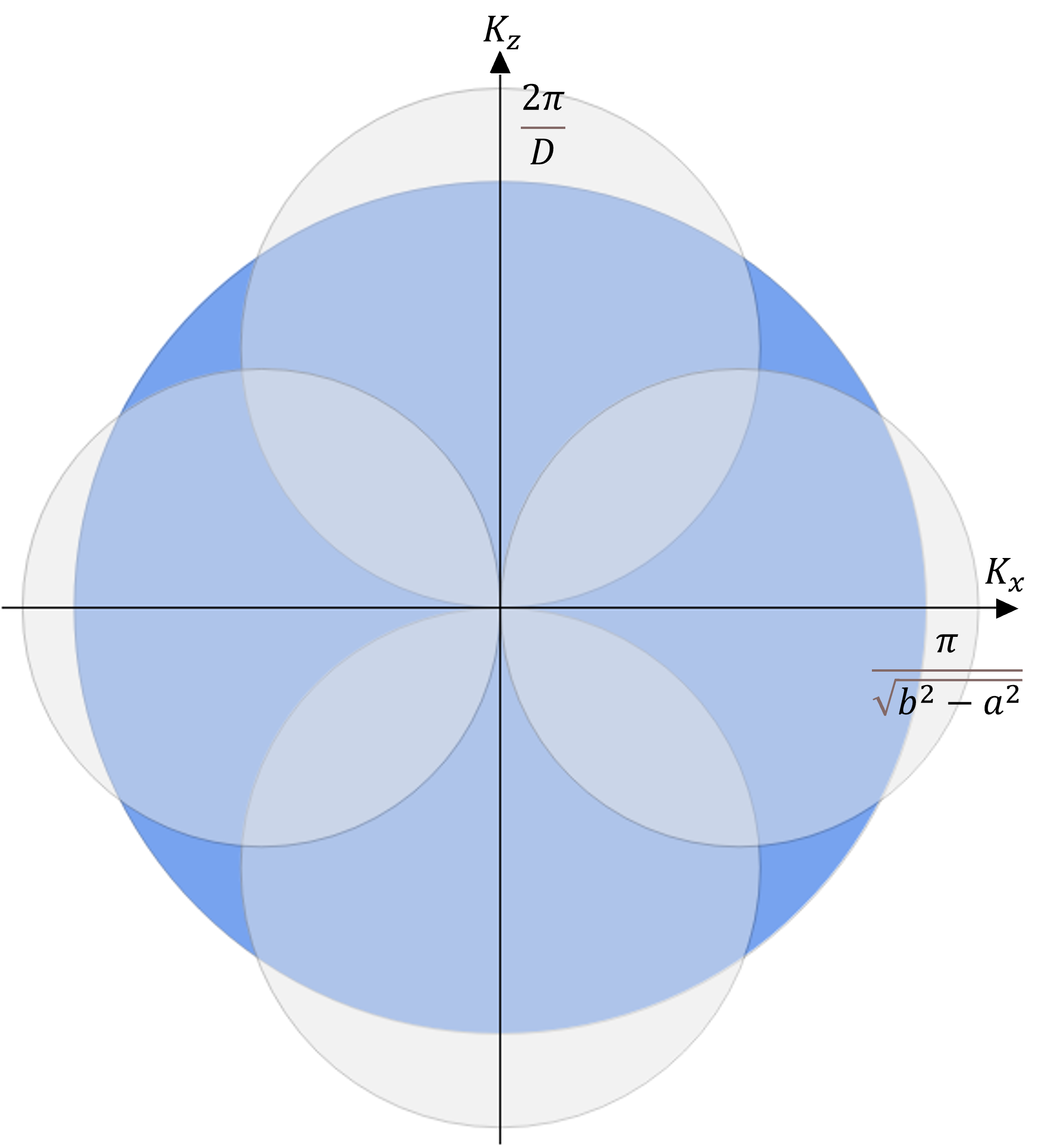}
    \caption{Projection of the 3D rendering of Fig. \ref{fig:rendering} in the $k_x-k_z$ space. In this case, both diameters that define the hole pockets of forbidden states are greater than the Fermi radius.}
    \label{fig:proj3}
\end{figure}

In Fig. \ref{fig:proj3} the corresponding situation is presented as a 2D projection on the $k_z$-$k_x$ plane.
\\

The fundamental homotopy group is therefore $\mathbb{Z}_6$.

In the following, to simplify some calculations we will suppose that the following inequality holds throughout:
\begin{equation}
    k_{F} > \frac{2\pi}{\sqrt{D^2+\mathcal{L}^2}}.
\end{equation}
This condition will define the range of applicability of our model.
As we are going to show in the next section, the quantity on the r.h.s. of the above inequality defines the radius of the spherical cap, in reciprocal space, resulting from the intersection between the Fermi sphere and the sphere of forbidden states (hole pocket).

\section{Computing Electron Density of States, Fermi Energy, and Critical Temperature for Confined Superconductors Using the Bardeen–Cooper–Schrieffer Gap Equation}

The goal of this section is to calculate the electron density of states (DOS), the Fermi energy, and the critical temperature for the quantum ring model introduced in the previous section and, where possible, compare our results with those obtained for thin films in \cite{Travaglino2023}.

To achieve this, we start by recalling the definition of the DOS:

\begin{equation}
    g(\epsilon)=  \diff{}{\epsilon}N(\epsilon'<\epsilon).\label{DOS2}
\end{equation}
This calculation becomes straightforward once the volume of occupied states in k-space, $V_k$, is determined as a function of the geometrical parameters. The number of occupied states with $k'<k$ is given by: 
\begin{equation}
    N(k'<k)=\frac{V}{2\pi^3} V_k.
\end{equation}

By transforming the variable from $k$ to $\epsilon$ using the dispersion relation for free electrons, we obtain $N(\epsilon'<\epsilon)$,  from which the DOS can be readily evaluated.
The complete calculation of  $V_k$ for this model is straightforward and presented in the Appendix.


If the confinement in the xy-plane is dominant (with respect to the confinement along the z-axis), we have that  $D>\mathcal{L}$.
Then the DOS is given by :
\begin{equation}
\label{e_density1}
   g(\epsilon) = \begin{cases}
        \frac{g_s V}{(2\pi)^2}\frac{(2m)^{\frac{3}{2}}}{\hbar^3}\sqrt{\epsilon} \quad \text{if} \quad \frac{\hbar^2}{2m}(\frac{2\pi}{\mathcal{L}})^2 < \epsilon < \epsilon_F,\\
        \frac{g_s V}{2\pi^3}\frac{\mathcal{L} m^2}{\hbar^4}\epsilon\quad \text{if}  
        \quad  \frac{\hbar^2}{2m}(\frac{2\pi}{D})^2 < \epsilon < \frac{\hbar^2}{2m}(\frac{2\pi}{\mathcal{L}})^2,\\
        \frac{g_s V}{2\pi^3}\frac{ m^2\epsilon}{\hbar^4}({D}+{\mathcal{L}})\quad \text{if} \\ \quad \quad \frac{\hbar^2}{2m}k_{cap}^2<\epsilon < \frac{\hbar^2}{2m}(\frac{2\pi}{D})^2.
    \end{cases}
\end{equation}
In the opposite case, when the confinement along the z-axis is dominant $D<\mathcal{L}$, the DOS is given by 
\begin{equation}
\label{e_density2}
   g(\epsilon) = \begin{cases}
        g_s\frac{V}{(2\pi)^2}\frac{(2m)^{\frac{3}{2}}}{\hbar^3}\sqrt{\epsilon} \quad \text{if} \quad \frac{\hbar^2}{2m}(\frac{2\pi}{D})^2 < \epsilon < \epsilon_F,\\
          g_s\frac{V}{2\pi^3}\frac{ m^2D}{\hbar^4}\epsilon\quad \text{if} \quad  \frac{\hbar^2}{2m}(\frac{2\pi}{\mathcal{L}})^2< \epsilon <  \frac{\hbar^2}{2m}(\frac{2\pi}{D})^2, \\
        g_s\frac{V}{2\pi^3}\frac{ m^2\epsilon}{\hbar^4}({D}+\mathcal{L})\quad \text{if} \\ \quad \quad
        \frac{\hbar^2}{2m}k_{cap}^2<\epsilon < \frac{\hbar^2}{2m}(\frac{2\pi}{\mathcal{L}})^2.
    \end{cases}
\end{equation}
Because the computation of the volume of the hole pockets in the regime $0<k_F<k_{cap}$ cannot be performed analytically (because it involves computing the intersection volume between three spheres), and is also less relevant for experimental applications, we do not compute the DOS in that regime. We will see later that, especially when we consider $k_{cap} \ll k_F$, this will be enough to compute a good approximation of the Fermi energy.

\begin{figure*}[ht]
    \centering
\includegraphics[width=0.8\textwidth]{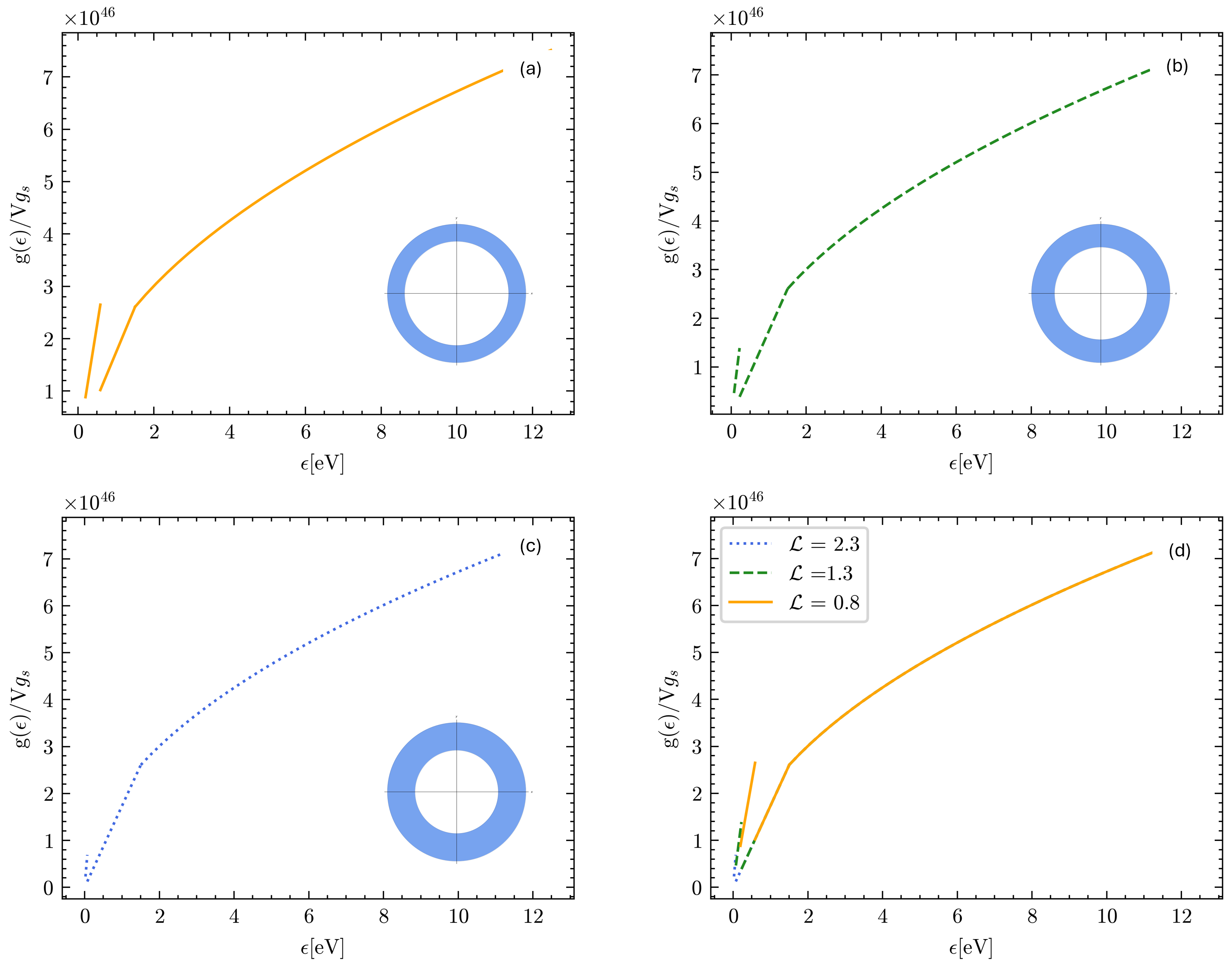}
    \caption{Electron density of states (DOS) plotted for three different choices of the confinement parameter $\mathcal{L}/2$ in the horizontal xy-plane. The confinement decreases from (a) to (b) to (c).
    The confinement along the z-axis was kept fixed as $D = 10^{-9}$ m. In (a) $\mathcal{L}/2$ was set to $0.8 \times 10^{-9}$ m. In (b) $\mathcal{L}/2$ was set to $1.3 \times 10^{-9}$ m. In (c) $\mathcal{L}/2$ was set to $2.3 \times 10^{-9}$ m. Panel (d) represents the overlap of the graphs (a), (b), (c).}
    \label{fig1}
\end{figure*}

\begin{figure*}[ht]
    \centering    \includegraphics[width=0.8\textwidth]{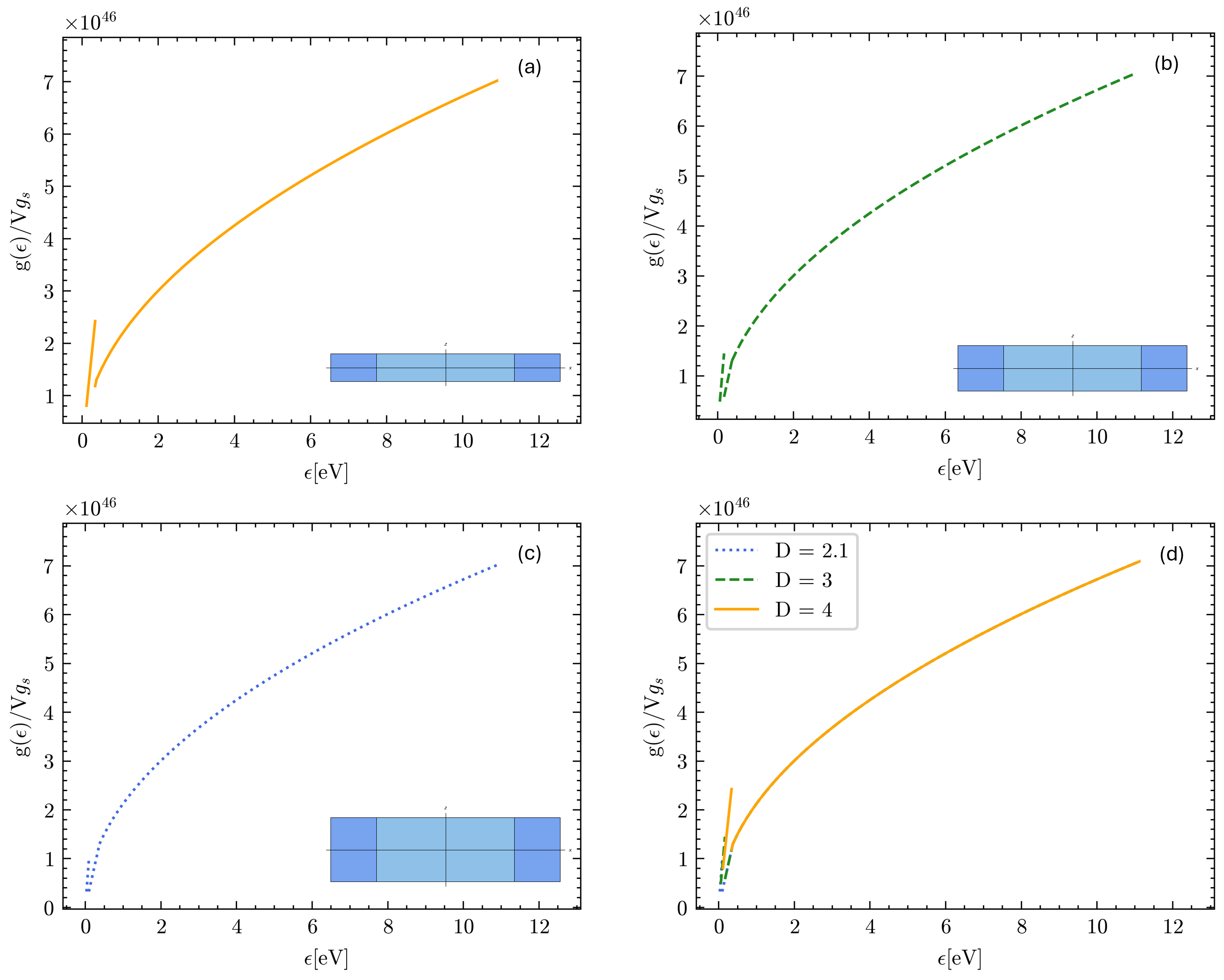}
    \caption{Electron density of states (DOS) plotted for three different choices of the confinement parameter $D$. The confinement in the xy-plane was kept fixed as $\mathcal{L}/2 = 10^{-9}$ m. In (a) $D$ was set to $2.1 \times 10^{-9}$ m. In (b) $D$ was set to $3 \times 10^{-9}$ m. In (c) $D$ was set to $4 \times 10^{-9}$. Panel (d) represents the overlap of the graphs (a), (b), (c).}
    \label{fig2}
\end{figure*}

In the following pictures, we present the DOS calculated according to the above formulae, as a function of the geometric confinement parameters in the xy plane, Fig. \ref{fig1}, and along the z-axis, Fig. \ref{fig2}.


We can see the signatures in the DOS of the two types of topological transition: 
the first kind is a transition from the bulk Fermi sphere, with homotopy group 0, to a topologically non-trivial surface with homotopy group $\mathbb{Z}$, which occurs when we transition from $\frac{\hbar^2}{2m}(\frac{2\pi}{\mathcal{L}})^2 < \epsilon < \epsilon_F$ to $\frac{\hbar^2}{2m}(\frac{2\pi}{D})^2 < \epsilon < \frac{\hbar^2}{2m}(\frac{2\pi}{\mathcal{L}})^2$ in Eq. \eqref{e_density1} (or, analogously, from $\frac{\hbar^2}{2m}(\frac{2\pi}{D})^2 < \epsilon < \epsilon_F$ to $\frac{\hbar^2}{2m}(\frac{2\pi}{\mathcal{L}})^2< \epsilon <  \frac{\hbar^2}{2m}(\frac{2\pi}{D})^2$ in Eq. \eqref{e_density2}).
The transition coincides with the change from the usual square-root behavior of the Fermi DOS (as for the bulk materials) to the linear in energy DOS predicted by the confinement theory \cite{Trachenko_2023}.
The DOS is continuous but not differentiable at the point of junction between the two topological regimes because there is a kink (qualitatively similar to what was found for thin films in Ref. \cite{Travaglino2023}). This happens when the Fermi level intersects two, out of four, hole pockets that define the states forbidden by the confinement (Fig. \ref{fig:proj2}). 

The second topological transition is a transition from a Fermi surface with homotopy group $\mathbb{Z}$, with two hole pockets intersecting the Fermi level, to a Fermi surface with homotopy group $\mathbb{Z}_6$, i.e. with all the four hole pockets intersecting the Fermi level (Fig. \ref{fig:proj3}).
This transition shows up as a discontinuity of the first kind in the DOS that occurs when we transition from  $\frac{\hbar^2}{2m}(\frac{2\pi}{D})^2 < \epsilon < \frac{\hbar^2}{2m}(\frac{2\pi}{\mathcal{L}})^2$ to $ \frac{\hbar^2}{2m}k_{cap}^2<\epsilon < \frac{\hbar^2}{2m}(\frac{2\pi}{D})^2$ in Eq.\eqref{e_density1} (or, analogously, form $\frac{\hbar^2}{2m}(\frac{2\pi}{\mathcal{L}})^2< \epsilon <  \frac{\hbar^2}{2m}(\frac{2\pi}{D})^2$ to $\frac{\hbar^2}{2m}k_{cap}^2<\epsilon < \frac{\hbar^2}{2m}(\frac{2\pi}{\mathcal{L}})^2$ in Eq.\eqref{e_density2}). The DOS is no longer continuous and it shows a jump at the junction of the two regimes. The jump separates two regimes that are both linear in the electron energy $\epsilon$.
This is because the Fermi level, $\epsilon_F$, switches from intersecting two of the hole pockets to intersecting all four of the hole pockets which are the forbidden states.

\hfill

Using the explicit expressions for the electron DOS obtained in the previous section, we can derive formulae for the Fermi energy $\epsilon_F$ of the quantum ring as a function of its geometric parameters $D$, $a$ and $b$. We will first consider the situation where the confinement in the xy-plane of the ring is stronger than the confinement along the axis of revolution (z-axis).

Because $N$ is the total number of electrons in the system, by definition of Fermi energy we get (at $T=0$):
\begin{equation}
    N = \int_0 ^{\epsilon_F} g_{s} g(\epsilon) d\epsilon,
\end{equation}
where $g_{s}$ accounts for spin degeneracy, and it is $g_s =2$ for electrons.

Furthermore, we define:
\begin{equation}
    \epsilon^* = \frac{\hbar^2}{2m}\left(\frac{2\pi}{D}\right)^2
\end{equation}
and \
\begin{equation}
    \epsilon^{**} = \frac{\hbar^2}{2m}\left(\frac{2\pi}{\mathcal{L}}\right)^2.
\end{equation}
Furthermore, for experimentally realizable confinement conditions, $k_{cap} < k_F$ holds. If this condition is violated the confinement would be extreme, corresponding to the values of the geometric parameters $D$ and $\mathcal{L}$  of the order of just a couple of Angstroms.

We first consider the case where the confinement in the xy-plane is dominant compared to the confinement along the z-axis.
The DOS is given by  Eq. \eqref{e_density1}. 
In this scenario, $\epsilon_F > \epsilon^{**} > \epsilon^{*}$ and the total number of electrons can be approximated as follows:
\begin{equation}
\begin{split}
     N&\simeq  \int_0 ^{\epsilon^{*}}  \frac{V}{\pi^3}\frac{ m^2\epsilon}{\hbar^4}({D}+{\mathcal{L}})d\epsilon +  \\
     & +\int_{\epsilon^{*}} ^{\epsilon^{**}} \frac{V}{\pi^3}\frac{\mathcal{L} m^2}{\hbar^4}\epsilon  d\epsilon+\int_{\epsilon^{**}} ^{\epsilon_F} 2 \frac{V (2m)^{3/2}}{(2\pi)^2 \hbar^3} \epsilon^{1/2} d\epsilon.
\end{split}
\end{equation}
This is true because $\int_0 ^{\epsilon_{cap}}g_sg(\epsilon)d\epsilon$ is small compared to the other integrals, where $\epsilon_{cap} = \frac{\hbar^2 k^2_{cap}}{2m}$.

The result gives
\begin{equation}
 \frac{N}{V} = \frac{4}{3}\frac{(2m)^{3/2}}{(2\pi)^2\hbar^3} \epsilon_F^{3/2} -\frac{2}{3} \frac{\pi}{(\mathcal{L})^3} + \frac{2\pi}{D^3}.
\end{equation}
Then the Fermi energy $\epsilon_F$, as a function of the geometrical parameters of the ring ($a$, $b$ and $D$) is 
\begin{equation}
\epsilon_F = \epsilon_F^{bulk}\Bigg[ 1+ \frac{2}{3}\frac{\pi}{\rho (\mathcal{L})^3} - \frac{2\pi}{\rho D^3}\Bigg]^{2/3}\label{Fermi1}
\end{equation}
where $\rho = N/V$ the density of the electron and $\epsilon_F^{bulk} = \frac{\hbar^2}{2m}(3\pi^2\rho)^{2/3}$.

This equation holds only if $\epsilon_F > \epsilon^{**}$. To determine the validity range of Eq. \eqref{Fermi1} as a function of the geometric parameters, we set $\epsilon_F = \epsilon^{**}$. This condition gives the following condition on the free-electrons density $\rho$: 
\begin{equation}
    \rho > 2\pi \left(\frac{1}{D^3} +  \frac{1}{\mathcal{L}^3}\right),
\end{equation}
and the Fermi energy is given by \eqref{Fermi1}. 
If, instead, $ \rho < \frac{2\pi}{D^3} +  \frac{2\pi}{\mathcal{L}^3}$, the Fermi energy
can be found through the new condition for $N$: 

\begin{equation}
\begin{split}
    \frac{N}{V}&=  \int_0 ^{\epsilon^{*}}  \frac{1}{\pi^3}\frac{ m^2\epsilon}{\hbar^4}({D}+{\mathcal{L}})d\epsilon\\& + \int_{\epsilon^{*}} ^{\epsilon_F} \frac{1}{\pi^3}\frac{\mathcal{L} m^2}{\hbar^4}\epsilon  d\epsilon  = \\
    & \frac{m^2\mathcal{L}}{2\pi^3\hbar^4}\epsilon_F^2+\frac{2\pi}{D^3},
\end{split}
\end{equation}

which results in 

\begin{equation}
    \epsilon_F = \frac{\hbar^2}{2m}\Bigg[ \frac{(2\pi)^3\rho}{\mathcal{L}}-\frac{(2\pi)^4}{\mathcal{L}D^3}\Bigg]^{1/2}.\label{Fermi2}
\end{equation}

For $\epsilon_F< \epsilon^*$ the approximation of $k_{cap}\ll k_F$ is no longer valid.


Analogously, for the conditions of Eq. \eqref{e_density2}, i.e. when the confinement along the z-axis is dominant compared to that in the xy-plane, we have:
\begin{equation}
\begin{split}
     \frac{N}{V}&\simeq  \int_0 ^{\epsilon^{**}}  \frac{ m^2\epsilon}{\pi^3\hbar^4}({D}+{\mathcal{L}})d\epsilon + \int_{\epsilon^{**}} ^{\epsilon^{*}} \frac{D m^2}{\pi^3\hbar^4}\epsilon  d\epsilon \\
     & +\int_{\epsilon^{*}} ^{\epsilon_F} 2 \frac{ (2m)^{3/2}}{(2\pi)^2 \hbar^3} \epsilon^{1/2} d\epsilon = \\
     &=\frac{4}{3}\frac{(2m)^{3/2}}{(2\pi)^2\hbar^3} \epsilon_F^{3/2} -\frac{2}{3} \frac{\pi}{D^3} + \frac{2\pi}{\mathcal{L}^3},
\end{split}
\end{equation}
and the corresponding Fermi energy is given by: 
\begin{equation}
\epsilon_F = \epsilon_F^{bulk}\Bigg[ 1+ \frac{2}{3}\frac{\pi}{\rho D^3} - \frac{2\pi}{\rho \mathcal{L}^3}\Bigg]^{2/3}. \label{Fermi3}
\end{equation}
This holds if $\epsilon_F > \epsilon^*$, or equivalently if:
\begin{equation}
     \rho > 2\pi \left(\frac{1}{D^3} +  \frac{1}{\mathcal{L}^3}\right).
\end{equation}

Otherwise, in the case of $\epsilon_F < \epsilon^*$, or $ \rho < \frac{2\pi}{D^3} +  \frac{2\pi}{\mathcal{L}^3}$, we have that 
\begin{equation}
\begin{split}
     \frac{N}{V}&\simeq  \int_0 ^{\epsilon^{**}}  \frac{  m^2\epsilon}{\pi^3 \hbar^4}({D}+{\mathcal{L}})d\epsilon + \int_{\epsilon^{**}} ^{\epsilon_F} \frac{D m^2}{\pi^3\hbar^4}\epsilon  d\epsilon\\
     &= \frac{(2m)^2D}{(2\pi)^3\hbar^4}\epsilon_F^2+\frac{2\pi}{\mathcal{L}^3}
\end{split}
\end{equation}
and the Fermi energy is given by:
\begin{equation}
    \epsilon_F = \frac{\hbar^2}{2m}\Bigg[ \frac{(2\pi)^3\rho}{D}-\frac{(2\pi)^4}{D\mathcal{L}^3}\Bigg]^{1/2}.\label{Fermi4}
\end{equation}

Equations \eqref{Fermi1}, \eqref{Fermi2}, \eqref{Fermi3}, \eqref{Fermi4} are among the most important results of this paper. They provide analytical expressions for the Fermi energy of metallic nanorings as a function of the ring geometric parameters, $a$ $b$ and $D$, and of the free-electron density of the material, $\rho$, in two regimes: Eqs. \eqref{Fermi1} and \eqref{Fermi2} are valid when the confinement in the xy-plane of the ring is stronger than the confinement along the vertical z-axis, and Eqs. \eqref{Fermi3} and \eqref{Fermi4} in the opposite case.

\hfill

Using the quantum confinement model outlined earlier, we can now analytically incorporate the effects of confinement into the BCS theory of superconductivity. Our last goal is to determine the critical temperature $T_c$ as a function of the confinement parameters $D$ and $\mathcal{L}$. In this context, we will use the mean-field weak-coupling approximation, to express the attractive phonon-mediated potential responsible for Cooper pairing $U_{\overrightarrow{k} \overrightarrow{k'}}$; for instance, we have \cite{Kittel,cohen}:
\begin{equation}
    U_{\overrightarrow{k} \overrightarrow{k'}} = \begin{cases}
        -U \quad \text{if} \quad |\epsilon-\epsilon_F|<\epsilon_D\\
        0 \quad\text{otherwise}
    \end{cases}
\end{equation}
where $\epsilon_D = \hbar \omega_D$ is the Debye (phonon) energy and $\omega_D$ is the Debye (phonon) frequency.

The energy gap at the ground state $\Delta$ depends on the density of states (DOS) at the Fermi level, which is influenced by quantum confinement effects, as derived in the previous sections. Consequently, the nanoring confinement alters the form of the energy gap that, in the weak-coupling limit, can be written as \cite{cohen}: 
\begin{equation}
    \Delta = 2\epsilon_D \exp\Bigg(-\frac{1}{Ug(\epsilon_F)}\Bigg).
\end{equation}

The critical superconducting temperature $T_c$ is related to the energy gap via $k_B T_c = \frac{2\Delta}{3.52}$, and is given by \cite{cohen}
    \begin{equation}
    T_c = \frac{4\epsilon_D}{3.52k_b} \exp\Bigg(-\frac{1}{Ug(\epsilon_F)}\Bigg).
\end{equation}

It is worth noting that, while the confinement changes the value of the Fermi energy, it does not affect the Debye energy which is a property
controlled solely by the atomic-scale structure and by the bonding physics of the specific material.

In the following, we will derive the form of the critical temperature $T_c$ for the metallic nanoring as a function of the ring geometric parameters, $D$, $a$, and $b$.

Let us suppose, without loss of generality, that $D>\mathcal{L}$.

In the regime for which $\epsilon_F>\epsilon^{**}$, the DOS at the Fermi energy is given by
\begin{equation}
    g(\epsilon_F) = g_{bulk}(\epsilon_F)\Bigg( 1+\frac{2}{3}\frac{\pi}{\rho \mathcal{L}^3}-\frac{2\pi}{\rho D^3}\Bigg)^{1/3}
\end{equation}
and, therefore, the critical temperature is 
\begin{equation}
    T_c = \frac{4\epsilon_D}{3.52k_B}\exp\Bigg[-\frac{1}{U g_{bulk}(\epsilon_F)\Big( 1+\frac{2}{3}\frac{\pi}{\rho \mathcal{L}^3}-\frac{2\pi}{\rho D^3}\Big)^{1/3}}\Bigg].
    \label{TC1}
\end{equation}
The corrections due to nanoring confinement indicate that the critical temperature $T_c$ increases as the values of the geometrical parameters decrease. Additionally, these corrections have a significant effect only if the inverse cube of the geometrical parameter is comparable to the free electron density in the material. Therefore, the superconducting critical temperature $T_c$ of the nanorings is predicted to be higher than that of the bulk superconductors. This increase remains relatively small until the geometrical parameters reach sufficiently small values, at which point the rise in $T_c$ compared to the bulk value becomes significant. However, if the parameters become too small, some of the mathematical approximations used in our derivation in the previous sections no longer hold.
For instance, the above formula for the critical temperature $T_c$ holds only if the geometrical parameters satisfy the inequality:
\begin{equation}
     \rho > 2\pi \left(\frac{1}{D^3} +  \frac{1}{\mathcal{L}^3}\right).
\end{equation}

If this is not the case, we are in the regime  $\epsilon_F<\epsilon^{**}$, in which all the four forbidden-states (hole-pocket) spheres cross the Fermi level.  We
will still assume that the DOS can be approximated by using its value at the Fermi energy: this is justified by the fact that the corrected DOS is continuous at the critical point, thus it can be Taylor expanded around $\epsilon_F$ with sufficient precision.
Proceeding as above, we find
\begin{equation}
    g(\epsilon_F) = g_{bulk}(\epsilon_F)\Bigg[ 2\pi \mathcal{L} \rho - \frac{(2\pi)^2\mathcal{L}}{D^3}\Bigg]^{1/2}
\end{equation}
that leads to the following formula for the critical temperature
\begin{equation}
    T_c = \frac{4\epsilon_D}{3.52k_B}\exp\Bigg[-\frac{1}{U g_{bulk}(\epsilon_F)}\frac{(3\pi^2 \rho)^{1/3}}{\Big( 2\pi \mathcal{L} \rho - \frac{(2\pi)^2\mathcal{L}}{D^3}\Big)^{1/2}}\Bigg].
\end{equation}
In this regime, the dependence of the $T_c$ on the geometric parameters is significantly different with respect to the previous case; this happens because $T_c$ now increases as the value of $D$ decreases, but it decreases if the value of $\mathcal{L}$ increases.

\begin{figure*}[ht]
    \centering    \includegraphics[width=0.88\textwidth]{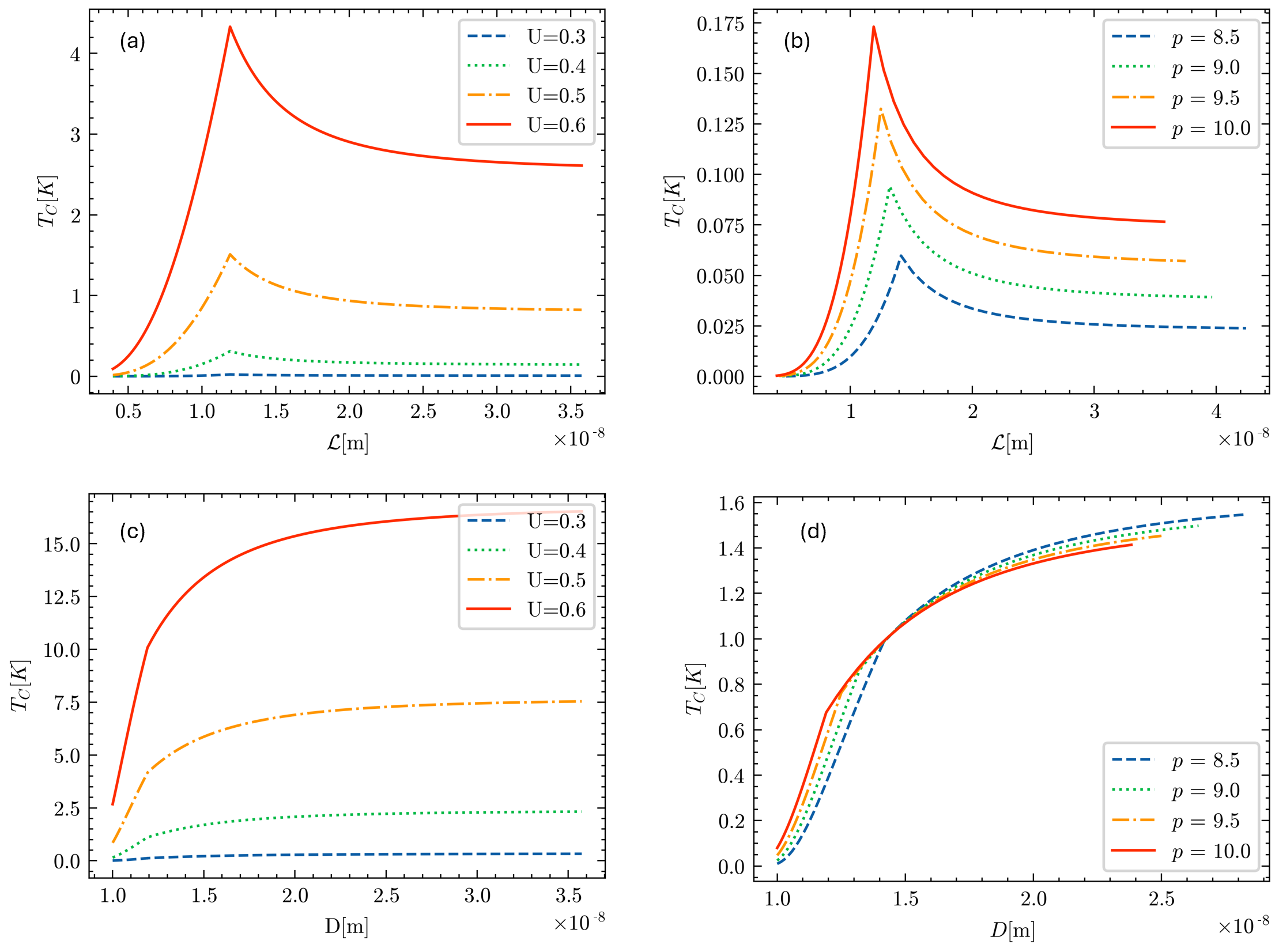}
    \caption{ Critical temperature $T_c$, in Kelvin, as a function of the geometrical parameters $\mathcal{L}$ and $D$, in meters. All the panels refer to the situation where the horizontal confinement is stronger than the vertical confinement, i.e. $D > \mathcal{L}$ (the opposite case $D < \mathcal{L}$ can be retrieved simply upon swapping $\mathcal{L}$ with $D$ in the abscissa of the plots). (a) Critical temperature $T_c$ vs $\mathcal{L}$ plotted for different values of $U$ while keeping $D$ fixed at $10^{-8}$ m. The density of states at the Fermi level is fixed $g(\epsilon_F) = 0.4$ eV$^{-1}$, with values of $U$ ranging from 0.3 to 0.6 eV.  (b) Critical temperature $T_c$ vs $D$ plotted for different values of $U$ while keeping $\mathcal{L}$ fixed at $10^{-8}$ m. The density of states at the Fermi level is fixed $g(\epsilon_F) = 0.4$ eV$^{-1}$, with values of $U$ ranging from 0.3 to 0.6 eV.  (c) Critical temperature $T_c$ vs $\mathcal{L}$ plotted as the density of electrons $\rho$ varies while keeping $U=0.37$ eV fixed. $\rho$ is expressed in units of $10^{24}$ m$^{-3}$.      (d) Critical temperature $T_c$ vs $D$ plotted as the density of electrons $\rho$ varies while keeping $U=0.37$ eV fixed. The free-carrier density $\rho$ is expressed in units of $10^{24}$ m$^{-3}$. }

\label{critical temperature}
\end{figure*}

Calculations based on the above theoretical model for $T_c$ as a function of the geometric parameters of the nanoring are shown in Fig. \ref{critical temperature}. In the top-left figure, we have fixed the value of
$D$ at $10^{-8}$ m and considered the free-electron density of nanoconfined aluminum as $\rho = 10^{25} $m$^{-3}$ \cite{Alcarriers}. The critical length scale $\mathcal{L}_c$ at which the topological transition from $\pi_1(S^2)=0$ to $\mathbb{Z}$ occurs is given by
\begin{equation}
    \mathcal{L}_c = \frac{(2\pi)^{1/3}}{(\rho-\frac{2\pi}{D^3})^{1/3}},
\end{equation}
which corresponds to a sharp maximum in the behavior of $T_c$ vs $\mathcal{L}$.

For $\mathcal{L}$ large compared to the cubic root of the free-carrier density $\rho$, the critical temperature is constant with $\mathcal{L}$ and then increases with further decreasing $\mathcal{L}$ up to the maximum, while for $\mathcal{L}< \mathcal{L}_c$ the critical temperature decreases with decreasing $\mathcal{L}$. 

To further illustrate the predicted behavior, we then let the Cooper pairing strength $U$ vary between 0.3 and 0.6 eV, in Fig. \ref{critical temperature}(a). 
The graph shows that increasing the pairing strength increases the critical temperature for a fixed $\mathcal{L}$ value, as expected.

In the bottom-left panel, Fig. \ref{critical temperature}(c), we still keep $D$ constant at $10^{-8}$ m and we also set $U$ fixed at 0.39 eV while we let the free-electron density $\rho$ vary between $8.5\times10^{24}$ m$^{-3}$ and $10^{25}$ m$^{-3}$. When the free-electron density increases, the value of $\mathcal{L}_c$ decreases as $\sim \rho^{-1/3}$. In addition, as the free-electron density $\rho$ increases, the maximum value of the critical temperature increases exponentially. Once again for $\mathcal{L}$ large compared to the cubic root of the free-electron density, the critical temperature is practically constant, while for $\mathcal{L}< \mathcal{L}_c$ the critical temperature decreases with further decreasing $\mathcal{L}$.

In the upper-right figure, Fig. \ref{critical temperature}(b), we have fixed $\mathcal{L}$ at $10^{-8}$ m and considered the free-electron density $\rho = 10^{25}$ m$^{-3}$. The critical value for $D$ at which the topological transition from $\pi_1(S^2)=0$ to $\mathbb{Z}$ occurs is then given by
\begin{equation}
    D_c = \frac{(2\pi)^{1/3}}{(\rho-\frac{2\pi}{\mathcal{L}^3})^{1/3}}.
\end{equation}
This value corresponds to a kink in the trend of $T_c$ vs $D$.

Also, in this case, we let the Cooper pairing strength $U$ vary between 0.3 and 0.6 eV. When $U$ increases, the value of the critical temperature also increases, as expected. We also see that for $D \rightarrow \mathcal{L}$, the value for the critical temperature tends to the value

\begin{equation}
    T_c = \frac{4\epsilon_D}{3.52k_B}\exp\Bigg[-\frac{1}{U g_{bulk}(\epsilon_F)\Big( 1 - \frac{4}{3}\frac{\pi}{\rho \mathcal{L}^3}\Big)^{1/3}}\Bigg]
\end{equation}   
that is smaller than the non-confined case (we shall return to this issue in the next section). 
The critical temperature as a function of $D$ is a monotonically increasing function, unlike the variation of $T_c$ with $\mathcal{L}$ examined above.

In the bottom-right figure, Fig. \ref{critical temperature}(d), we still keep $\mathcal{L}$ constant at $10^{-8}$ m and we also set $U$ at 0.37 eV while we let the free-electron density vary between $8.5\times10^{24}$ m$^{-3}$ and $10^{25}$ m$^{-3}$. When the free-electron density increases, the value of $D_c$ decreases as $\sim \rho^{-1/3}$.  

Until now, it has been assumed that $D>\mathcal{L}$. If, instead, $D<\mathcal{L}$, with steps analogous to the previous case, we get that the critical temperature is given by
\begin{equation}
    T_c = \frac{4\epsilon_D}{3.52k_B}\exp\Bigg[-\frac{1}{U g_{bulk}(\epsilon_F)\Big( 1+\frac{2}{3}\frac{\pi}{\rho D^3}-\frac{2\pi}{\rho \mathcal{L}^3}\Big)^{1/3}}\Bigg]\label{cross-2}
\end{equation}
if 
\begin{equation}
    3\pi^2 \rho > \frac{6\pi^3}{D^3} +  \frac{6\pi^3}{\mathcal{L}^3}.
\end{equation}
and by 
\begin{equation}
    T_c = \frac{4\epsilon_D}{3.52k_B}\exp\Bigg[-\frac{1}{U g_{bulk}(\epsilon_F)}\frac{(3\pi^2 \rho)^{1/3}}{\Big( 2\pi D \rho - \frac{(2\pi)^2D}{\mathcal{L}^3}\Big)^{1/2}}\Bigg]
\end{equation}
otherwise. So we notice that the form of the critical temperature is the same as in the previous case, except for swapping $D$ and $\mathcal{L}$. The qualitative trends for the critical temperature and the conclusions that can be drawn are also the same as in Fig. \ref{critical temperature}, except for swapping $D$ and $\mathcal{L}$ in the abscissa of the graphs.

It should be noted that the cusp at the peak (discontinuity of the first derivative) is due to
the approximations made in the previous sections. If the whole DOS (instead of the constant-DOS approximation) were considered in the solution of the gap equation, then a slight rounding effect would remove the cusp at the peak, as discussed in \cite{Travaglino2023}.

\hfill

Up to this point, we have divided our study into two cases, assuming that one of the two characteristic dimensions of the ring, either $D$ or $\mathcal{L}$, is larger than the other. In this section, we will analyze the special case where $D=\mathcal{L}$, as a limiting case of the results obtained above.

For clarity, we define a new parameter $L$ related to the previous parameters by the relationship $L \equiv D = \mathcal{L}$. This will be the only parameter on which the physical quantities (e.g. the $T_c$) depend.

The primary difference compared to the previous cases is that all four spheres defining the hole pockets have equal radii. As a result, the only possible topological transition is from the homotopy group 0 to $\mathbb{Z}_6$ because the system moves from a state where the Fermi sphere is not intersected by any of the four hole pockets to a state where all four hole pockets intersect the Fermi sphere. 

If we suppose that $k_{F} >\frac{2\pi}{L}$ holds, the geometry in momentum space is described by Fig. \ref{fig:proj1}. Otherwise, the geometry in momentum space is given by Fig. \ref{fig:proj3}.

The volume of the occupied states in k-space can then be calculated as follows:

\begin{itemize}
    \item \text{if} $k_F>\frac{2\pi}{L}$:\\
    $V_k = \frac{4\pi}{3}{k_F}^3 -4\frac{4\pi}{3}\big({\frac{\pi}{L}}\big)^3 +4V_{intersect} $
    \item \text{if} $k_F<\frac{2\pi}{L}$:\\
     $V_k = \frac{4\pi}{3}{k_F}^3 -4\frac{4\pi}{3}\big({\frac{\pi}{L}}\big)^3 +4V_{intersect} +4V_{outside}$
\end{itemize}
where, in the second equation, $V_{outside}$ is evaluated at $k=\frac{\pi}{L}$.
Also, we have assumed, as before, that $2k_{cap}<k_F$. Using these formulae, and following the same method applied in the previous cases, one can now calculate the density of states (DOS) for the special case $\mathcal{L}=D$.

Another, more straightforward, method to calculate $g(\epsilon)$ is by taking the limit $\mathcal{L} \rightarrow D$ in Eq. \eqref{e_density1}.  Note that because $\epsilon^{**}\rightarrow \epsilon^{*}$, there are only two regimes: the first one for $\epsilon$ in the range $\frac{\hbar^2}{2m}\big(\frac{2\pi}{L}\big)<\epsilon<\epsilon_F$ and the second one for $\frac{\hbar^2}{2m}k_{cap}^2<\epsilon<\frac{\hbar^2}{2m}\big(\frac{2\pi}{L}\big)$. 

Then, by substituting $D$ and $\mathcal{L}$ with $L$ we get:
\begin{equation}
\label{e_density_eq}
   g(\epsilon) = \begin{cases}
        \frac{g_s V}{(2\pi)^2}\frac{(2m)^{\frac{3}{2}}}{\hbar^3}\sqrt{\epsilon} \quad \text{if} \quad \epsilon^{\dagger} < \epsilon < \epsilon_F,\\
        \frac{g_s V}{\pi^3}\frac{ m^2 L}{\hbar^4}{\epsilon}\quad \text{if}  \quad \frac{\hbar^2}{2m}k_{cap}^2<\epsilon < \epsilon^{\dagger}.
    \end{cases}
\end{equation}
where $\epsilon^{\dagger}$ is defined as $\epsilon^{\dagger} = \frac{\hbar^2}{2m}(\frac{2\pi}{L})^2$.

Once again, we do not compute the DOS
in the regime $0<\epsilon<\frac{\hbar^2}{2m}k_{cap}^2$ due to complications in analytically calculating the volume of the holes pocket in that region of the momentum space.

\begin{figure}[h]
    \centering    \includegraphics[width=0.5\textwidth]{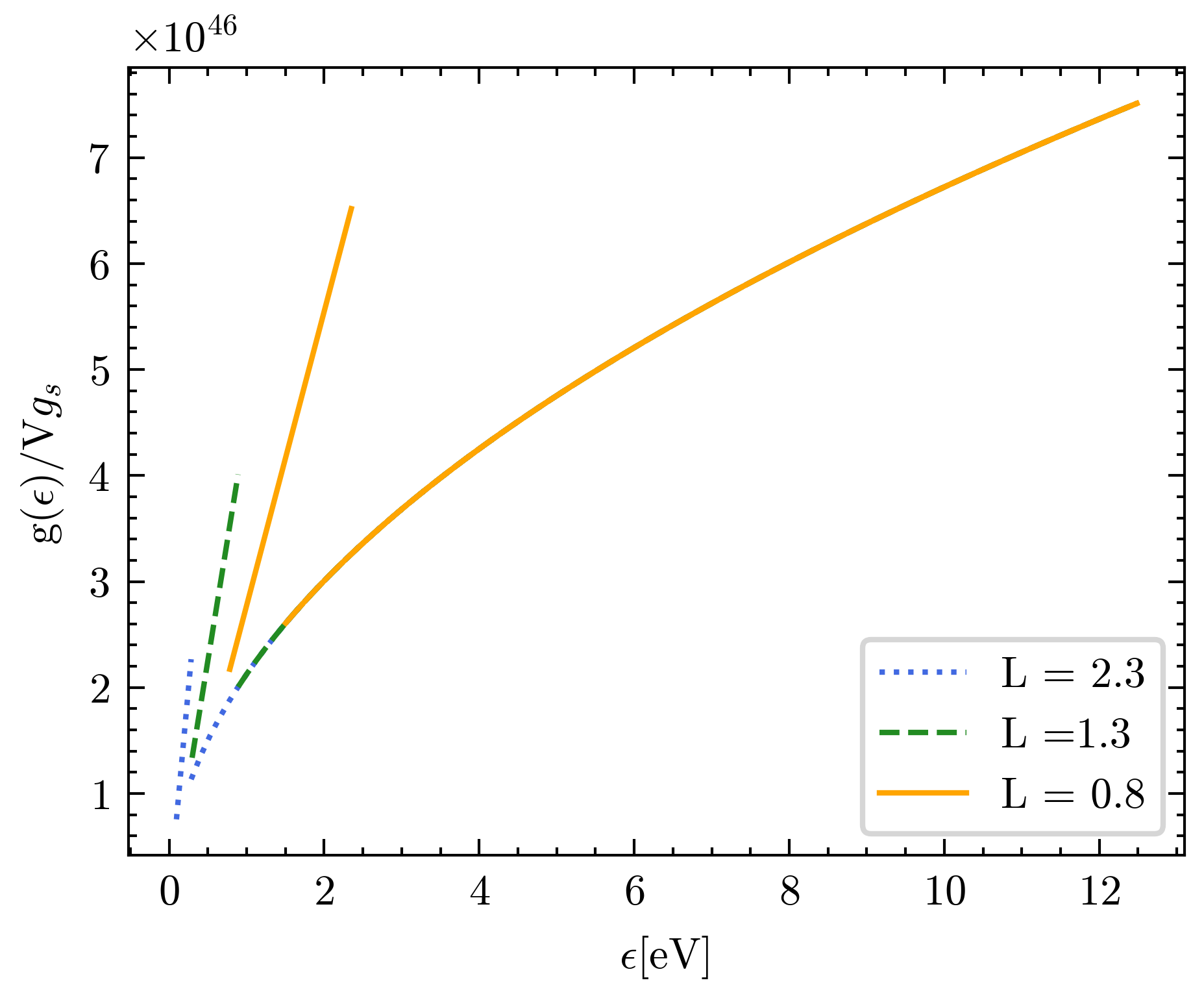}
    \caption{Electron density of states (DOS) plotted for three different choices of the confinement parameter $L=\mathcal{L}=D$, which is expressed in units of $10^{-9}$ m (see legend). 
    }

\label{energyDensityEqual}
\end{figure}

In Fig. \ref{energyDensityEqual} the profile of $g(\epsilon)$ as a function of $\epsilon$ is shown. Specifically, $g(\epsilon)$ has been plotted for different values of $L$ spanning from $0.8 \times10^{-9}$ m to $2.3 \times10^{-9}$ m.

As expected, there is only one topological transition from the bulk Fermi sphere, with a homotopy group 0, to a topologically non-trivial surface with homotopy group $\mathbb{Z}_6$, which occurs when we transition from $\frac{\hbar^2}{2m}k_{cap}^2<\epsilon < \epsilon^{\dagger}$ to $\quad \epsilon^{\dagger} < \epsilon < \epsilon_F$, or analogously, when the Fermi level intersects all the four hole pockets that define the states forbidden by the confinement. 

The break visible in the plotted DOS curves in Fig. \ref{energyDensityEqual} is a first-order discontinuity characterized by a jump in $\epsilon^{\dagger}$, which separates the usual square-root behavior of the Fermi-gas DOS from the linear regime determined by confinement (compare with thin films, in \cite{Travaglino2023}). As the confinement length $L$ increases, the magnitude of the jump decreases and gets shifted to lower energy values, as expected for a less confined regime, while it becomes significant, and shifted to larger energy values when $L$ takes on smaller values.
Using the explicit expressions for the electron density of states (DOS), similarly to our previous approach, we can derive formulae for the Fermi energy as a function of the geometric parameter $L$. The result can be stated as follows: 

\begin{equation}
         \epsilon_F = \begin{cases}
         \epsilon_F^{bulk}\Big( 1- \frac{4}{3}\frac{\pi}{\rho L^3}\Big)^{2/3} \text{if} \quad L>L_{c} = \Big(\frac{4\pi}{\rho}\Big)^{1/3} \\
         \frac{\hbar^2}{2m} \Big(\frac{4\pi^3\rho}{L}\Big)^{1/2} \text{if}\quad L<L_{c}         
    \end{cases} 
\end{equation}

where $\epsilon_F^{bulk} = \frac{\hbar^2}{2m}(3\pi^2\rho)^{2/3}$.

The first equation is consistent with what was discussed in previous sections. Specifically, by substituting $L=D=\mathcal{L}$ in Eq. \eqref{Fermi1}, one can verify that the obtained result matches the previous findings.

This is not true for the second formula (the one valid for $L<L_c$). This discrepancy arises because, in that regime, the approximation $k_{cap}\ll k_F$ is no longer valid. 
As there is, now, only one geometric parameter $L$, it is possible to calculate the value of $L$, referred to as $L_{c}$, that corresponds to the topological transition marking the change from one Fermi energy formula to the other. When comparing the value of $L_{c}$ in this paper with the one found for thin films in \cite{Travaglino2023}, it appears that the value under the cubic root is exactly twice the value obtained in \cite{Travaglino2023}. This difference arises because our calculation also accounts for the horizontal in-plane confinement along the x-y axes.

We are now poised to delve into the computation of the critical temperature for the onset of superconductivity in the nanoring, in this special case.

\begin{figure*}[ht]
    \centering    \includegraphics[width=0.87\textwidth]{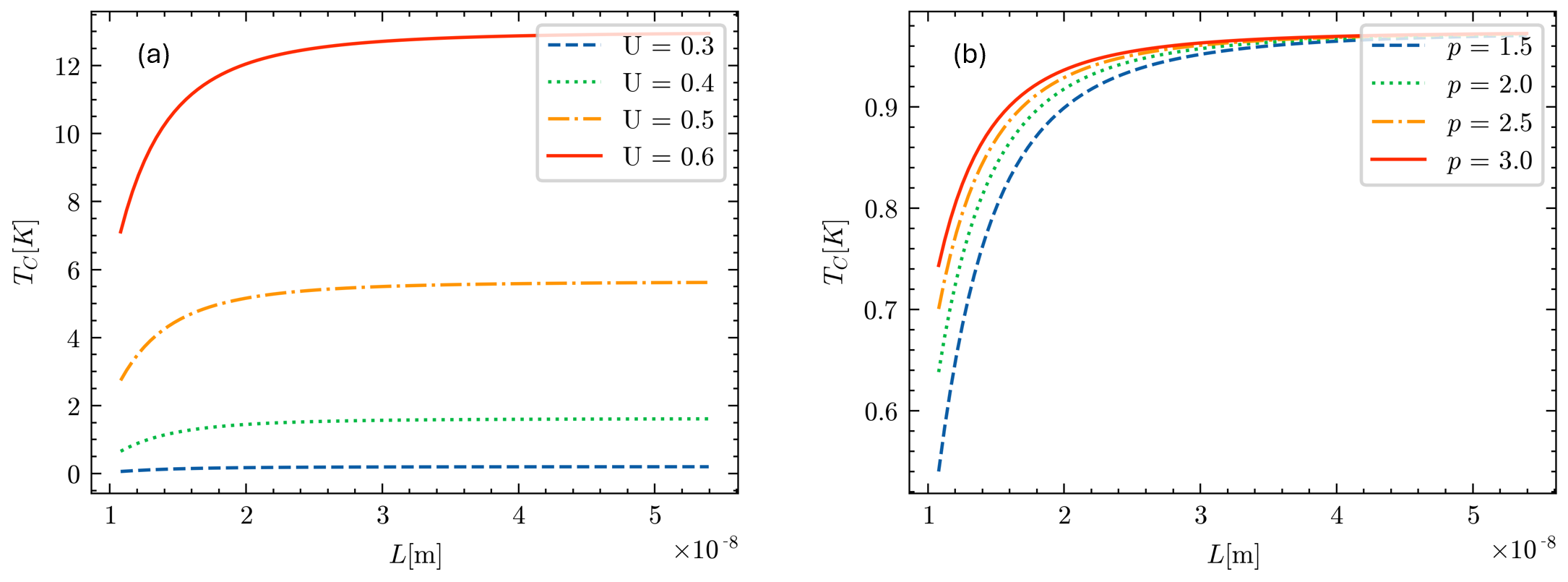}
    \caption{Critical temperature $T_c$, in Kelvin, as a function of the geometrical parameters $L$ in meters. (a) Critical temperature $T_c$ vs $L$ plotted for different values of $U$ that span from 0.3 to 0.6 eV, keeping the density fixed at $\rho = 10^{25}$ and the DOS at the Fermi level at $g(\epsilon_F) = 0.4$ eV$^{-1}$. The value of $L_c$ is $1.079 \times 10^{-8}$ m. (b) Critical temperature $T_c$ vs $L$ plotted for different values of $\rho$ that span from 1.5 to 3 in unit of $10^{25}$, keeping $U$ fixed at $3.7$ eV and the DOS at the Fermi level at $g(\epsilon_F) = 0.4$ eV$^{-1}$. The value of $L_c$ depend on $\rho$, and are $9.4 \times 10^{-9}$, $8.6 \times 10^{-9}$, $7.95 \times 10^{-9}$, $7.5 \times 10^{-9}$, in order of increasing $\rho$. }

\label{TCEqual}
\end{figure*}

In this case, the Fermi energy is correctly approximated only in the regime where $L>L_c$. Hence, we can calculate the critical temperature $T_c$ only within this regime. The result is:

\begin{equation}
    T_c = \frac{4\epsilon_D}{3.52k_B}\exp\Bigg[-\frac{1}{U g_{bulk}(\epsilon_F)\Big( 1 - \frac{4}{3}\frac{\pi}{\rho L^3}\Big)^{1/3}}\Bigg],
\end{equation}   
and the corresponding illustrative plots are reported in Fig. \ref{TCEqual}.
In the regime where $L>L_c$, the critical temperature $T_c$ is a strictly monotonically increasing function of the geometrical parameter $L$.  

In Fig. \ref{TCEqual}(a), the free-electron density of nanoconfined aluminum is kept fixed as $\rho = 10^{25}$ 
 m$^{-3}$, the critical temperature is plotted as a function of the geometrical parameter $L$ when the Cooper pairing strength $U$ varies between 0.3 and 0.6 eV.
As in the previous case, the graph demonstrates that increasing the pairing strength $U$ increases the critical temperature $T_c$ for a fixed $L$ value, as expected. For this set of curves, the $L_c$ is in the order of $10^{-9}-10^{-8}$ m. 

In Fig. \ref{TCEqual}(b), after fixing $U= 0.37$ eV, the critical temperature is plotted as a function of the geometrical parameter $L$ while allowing the free-electron density $\rho$ to vary between $1.5 \times10^{25}$ m$^{-3}$ and $3 \times10^{25}$ m$^{-3}$. As the free-electron density $\rho$ increases, the value of $L_c$ decreases approximately as $\rho^{-1/3}$, within a span from 7 to 9 $\times10^{-9}$ m$^{-3}$. It is also seen that the $T_c$ is a monotonically increasing function of $L$, and it increases as the free-electron density increases.

The regime $L> L_c$ is the only one for which we can obtain analytical solutions. It is expected that in correspondence of $L=L_c$ (not shown in Fig. \ref{TCEqual}) there will be a kink (discontinuity in the first derivative) similar to the one observed for the case of strong in-plane confinement in Fig. \ref{critical temperature}(b) and (d).

Finally, we focus on the existence and the behavior of the following transitions:
\begin{itemize}
    \item From the regime of Fig. \ref{critical temperature} (a) and (b) to the one of Fig. \ref{critical temperature}(c) and (d);
    \item From the regime $\mathcal{L}>D$ to the regime $\mathcal{L}<D$.\\
\end{itemize}
Of course, these two crossovers are essentially just the same crossover, and we can speak of a unique crossover from $D>\mathcal{L}$ to $D<\mathcal{L}$.

To this aim, it is sufficient to study the portion of the critical temperature that depends on the parameters of the ring, referred to as the correction function throughout the rest of the discussion. This geometry-dependent correction, for instance, is $1+\frac{2}{3}\frac{\pi}{\rho \mathcal{L}^3}-\frac{2\pi}{\rho D^3}$ for $D>\mathcal{L}$ (cf. Eq. \eqref{TC1})  and $1+\frac{2}{3}\frac{\pi}{\rho D^3}-\frac{2\pi}{\rho \mathcal{L}^3}$ for $D<\mathcal{L}$ (cf. Eq. \eqref{cross-2}).

Let us start with $D>\mathcal{L}$ so that the critical temperature is given by Eq. \eqref{TC1}. 

The critical temperature exhibits the behavior shown in Fig. \ref{critical temperature}(a) and (c) if the correction function is greater than 1, and the behavior shown in Fig. \ref{critical temperature}(b) and (d) if the correction function becomes less than 1. This has a simple yet meaningful explanation. Suppose the correction function is equal to 1. In that case, the critical temperature obtained is the same as in the non-confined (bulk) case. If the correction function is less than 1, then the critical temperature is lower than that in the non-confined case. Conversely, if the correction function is greater than 1, then the critical temperature is higher than that in the non-confined case. 

Let us discuss now how the critical temperature varies for values of the parameter $(\mathcal{L}, D) = (L, L\pm \delta)$, where $\delta$ is such that $\delta/L \ll 1 $.

We will focus on the range $\rho > \frac{2\pi}{L^3} + \frac{2\pi}{(L+\delta)^3}$, because only this regime is well-defined in the limit $\delta \rightarrow 0$. After Taylor-expanding the correction function in $\delta/L$, the critical temperature for small $\delta$ is well approximated by:
\begin{equation}
    T_c = \frac{4\epsilon_D}{3.52k_B}\exp\Bigg[-\frac{1}{U g_{bulk}(\epsilon_F)\Big( 1 - \frac{4}{3}\frac{\pi}{\rho {L}^3} + \frac{6\pi}{\rho L^4}\delta\Big)^{1/3}}\Bigg] 
\end{equation}  
if $D>\mathcal{L}$ and by 
\begin{equation}
    T_c = \frac{4\epsilon_D}{3.52k_B}\exp\Bigg[-\frac{1}{U g_{bulk}(\epsilon_F)\Big( 1 - \frac{4}{3}\frac{\pi}{\rho {L}^3} + \frac{2\pi}{\rho L^4}\delta\Big)^{1/3}}\Bigg] 
\end{equation}  
if $D<\mathcal{L}$.

We first notice that the correction function is smaller than one, so the expected behavior for the special case $\mathcal{L}=D=L$ will be qualitatively similar to the trend in Figs. \ref{critical temperature}(b) and (d) (i.e. $T_c$ monotonically increasing with $L$, with a kink at $L_c$). In addition, the critical temperature has a minimum value for $D=\mathcal{L}$, for reasons discussed above. 

Finally, to investigate the transition from the behavior depicted in Fig. \ref{critical temperature} (a) and (c) to the one shown in Fig. \ref{critical temperature} (b) and (d), we study the critical temperature for the values of the parameter $(\mathcal{L}, D) =(L, L+\epsilon) $, i.e. for different values of $\epsilon >0$. We notice that, because the assumption $D>\mathcal{L}$ holds, we can plot Eq. \eqref{TC1} slightly rewritten in the following form:
\begin{equation}
    T_c = \frac{4\epsilon_D}{3.52k_B}\exp\Bigg[-\frac{1}{U g_{bulk}(\epsilon_F)\Big( 1+\frac{2}{3}\frac{\pi}{\rho L^3}-\frac{2\pi}{\rho (L+\epsilon)^3}\Big)^{1/3}}\Bigg].
    \label{TC2}
\end{equation}
by restricting ourselves to the case where $L$ satisfies the inequality:
\begin{equation}
    \rho > \frac{2\pi}{L^3}+\frac{2\pi}{(L+\epsilon)^3}.
\end{equation}

\begin{figure}[ht]
   \centering    \includegraphics[width=0.5\textwidth]{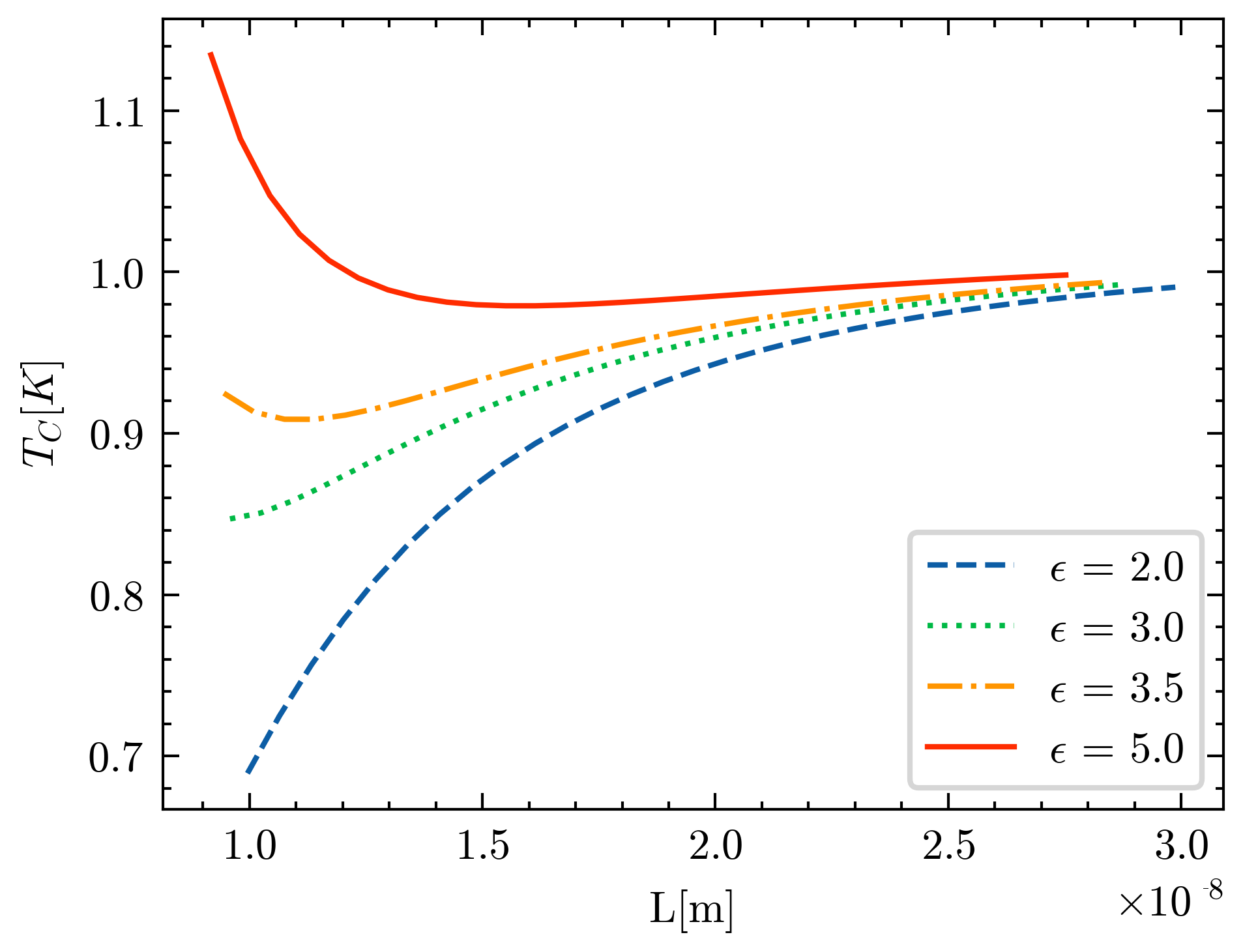}
    \caption{Critical temperature $T_c$, in Kelvin, as a function of the geometrical parameter $L$, in meters. The critical temperature $T_c$ vs $L$ is plotted while keeping $U=0.37$ eV, $\rho = 10^{25}$ m$^{-3}$ and the DOS at the Fermi level at $g(\epsilon_F) = 0.4$ eV$^{-1}$ as constant. The parameter $\epsilon$ in Eq. \eqref{TC2} varies between $2\times10^{-9}$ m and $5\times10^{-9}$ m, as detailed in the legend.}
\label{TCE_transition_curvature}
\end{figure}

The result is shown in Fig. \ref{TCE_transition_curvature}.

The transition from the monotonically increasing regime of $T_c$ vs $L$ to the one with the maximum, appears to be a smooth function of the parameter $\epsilon$. Specifically, for a fixed $L$, as $\epsilon$ increases, the negative component of the correction function decreases, leading to a rise in the critical temperature. This aligns with our previous findings: when $D = \mathcal{L}$, the critical temperature reaches a minimum. Conversely, as $\epsilon \rightarrow \infty$, we retrieve the case of no confinement along the z-axis. Here, the critical temperature has the largest values, and exhibits a behavior consistent with that described in Ref. \cite{Travaglino2023} for thin film confinement.

\section{Conclusions}
In summary, we have developed an exact analytical solution to compute the effect of confinement on the available volume of accessible states in momentum space for free electrons propagating in a confined metallic sample with the toroid geometry sketched in Fig. \ref{fig:real-space}, also referred to as nanoring or quantum ring.

For a gas of free fermions, the available volume
in momentum space is just the Fermi k-sphere, but, in the presence of confinement, this is no longer true, and four spheres of forbidden states (\emph{hole pockets}) develop inside the Fermi sphere. This is valid as long as the confinement is not below the nanometer scale, which is, anyway, difficult to achieve experimentally. Their
size grows with the shrinking of the geometric parameters of the ring, i.e. the ring thickness $D$ and the difference of the two radii-squared $b^2-a^2$, as illustrated in Fig. \ref{fig:rendering}.
Two hole-pocket spheres in k-space are associated with confinement in the plane of the toroid, i.e. they grow upon shrinking $b^2-a^2$, while the other two spheres are associated with confinement along the vertical direction z, i.e. they grow upon shrinking the value of $D$.

Two distinct topological transitions in the Fermi surface are predicted, upon increasing the confinement, i.e. upon decreasing the values of $D$ and of $b^2-a^2$. The first transition occurs when the confinement along the vertical direction z is dominant with respect to the confinement in the xy plane. In this case, a topological transition occurs when the two hole-pocket spheres grow across the Fermi sphere along $k_z$. As the Fermi surface is "disrupted" in this way, we transition from the standard Fermi surface for the free electron gas, with the (trivial) fundamental homotopy group $\pi_1(S^2)=0$, to a topologically non-trivial Fermi surface with fundamental homotopy group $\mathbb{Z}$ (as in the case of thin films \cite{Travaglino2023}). This happens while the two hole-pocket spheres associated with xy-confinement remain within the Fermi sphere.
As these latter two hole-pocket spheres also grow across the Fermi sphere, and "disrupt" it, we then transition from a Fermi surface with fundamental homotopy group $\mathbb{Z}$ to a Fermi surface with fundamental homotopy group $\mathbb{Z}_6$.

This geometric distortion of the available momentum space has profound consequences on the DOS and, under certain hypotheses, we were able to analytically evaluate the DOS for the nanoring confinement. The results have been graphically plotted in Fig. \ref{fig1} and Fig. \ref{fig2}.
The two topological transitions in the Fermi surface show up in the electron DOS, and appear as a kink and as a jump, respectively. The first transition, from $\pi_1(S^2)=0$ to $\mathbb{Z}$, shows up as a kink in the DOS, marking the change from the standard square-root behavior of the free-electron DOS, to the linear-in-energy behavior due to confinement (qualitatively similar to what happens in thin films \cite{Travaglino2023}). The second topological transition is evident as a jump in the DOS which separates two regimes that are both linear in energy.

Based on these results, analytical closed-form expressions are derived for the Fermi energy as a function of the geometric parameters of the nanoring, $\mathcal{L}=\sqrt{b^2-a^2}$ and $D$.

The obtained results for the electronic DOS and for the Fermi energy as a function of the nanoring geometric parameter values are subsequently implemented in the BCS theory of superconductivity, in the weak-coupling limit valid e.g. for aluminum nanorings. 

In particular, when the horizontal (in-plane) confinement is dominant with respect to the vertical confinement ($\mathcal{L}>D$), the mathematical theory predicts that the superconducting critical temperature $T_c$ varies non-monotonically as a function of the in-plane ring confinement, measured by the parameter $\mathcal{L}=\sqrt{b^2-a^2}$. Indeed, the $T_c$ is predicted to show a maximum at a critical in-plane confinement value $\mathcal{L}_c$, which depends solely on the free-electron concentration in the sample and on the other geometric parameter $D$. The maximum corresponds to the topological transition in the Fermi surface from the Fermi sphere to the distorted surface with fundamental homotopy group $\mathbb{Z}$. The $T_c$ is also predicted to increase monotonically with the vertical confinement parameter $D$, with a kink (discontinuity in the first derivative) that depends on the free carriers density $\rho$. 
If, instead, it is the vertical confinement that dominates over the horizontal one, the situation is reversed: there is a maximum in the $T_c$ as a function of the vertical confinement parameter $D$, and a monotonically increasing trend with a kink as a function of $\mathcal{L}$.

In the special case of a square toroid, $\mathcal{L}=D=L$, there is only one topological transition, from $\pi_1(S^2)=0$ to $\mathbb{Z}_6$. The critical temperature does not present a maximum, and the $T_c$ monotonically increases upon increasing the confining length scale $L$, as long as one remains in the regime $L > L_c =(4\pi/\rho)^{1/3}$ (which is the only regime analytically tractable within the proposed theory). A kink is expected to occur at $L_c$.
By studying the vicinity of the special point $\mathcal{L}=D$ in parameter space, we have shown that there is a crossover from the monotonic behavior to the non-monotonic one with the maximum, as $D$ becomes gradually larger than $\mathcal{L}$.

These results open up new directions of research involving quantum nanorings, in particular, with reference to the on-demand design of superconductivity in quantum rings. These new directions include the possibility of tuning the superconducting critical temperature by controlling the geometric parameters of the ring. The geometric parameters control the DOS and the Fermi energy as per the formulae derived in the present paper, and hence the superconducting critical temperature $T_c$. One of the main methods to determine the critical temperature in superconducting quantum rings  consists in applying the "Little's fit" \cite{Bezryadin2008, Tinkham2002} to the experimentally detected resistance($R$)-versus-temperature($T$) transition (see, e.g., Fig. 1b in Ref. \cite{Papari2019}): 
\begin{equation}
   \frac{R(T)}{R(T_c)}=\exp\Bigg[-\frac{\sqrt{6}\Phi_0 I_c(0)}{k_B T}\Big( 1-\frac{T}{T_c}\Big)^{3/2}\Bigg],
    \label{Arrhenius}
\end{equation}
where $R(T_c)$ is the normal resistance achieved at the critical temperature, $\Phi_0$ is the magnetic flux quantum, the critical current at zero temperature $I_c(0)=j_c(0)D(b-a)$ is a product of the critical current density $j_c(0)$ with the cross-sectional area of the ring in the plane passing through the ring's axis. The width of the ring is related to the parameter $\mathcal{L}$ by the formula $b-a=a+\sqrt{a^2+(\mathcal{L}/2)^2}$. 

According to the findings of the present paper, the ring's critical temperature is a function of the geometric parameters $\mathcal{L}$ and $D$, the relative resistance mediated by the exponential function of $T_c$ and the critical current density in Eq. (\ref{Arrhenius}) should manifest even a more prominent dependence on $\mathcal{L}$ and $D$ in the experimental measurements. 

\begin{figure*}[ht]
    \centering    \includegraphics[width=0.87\textwidth]{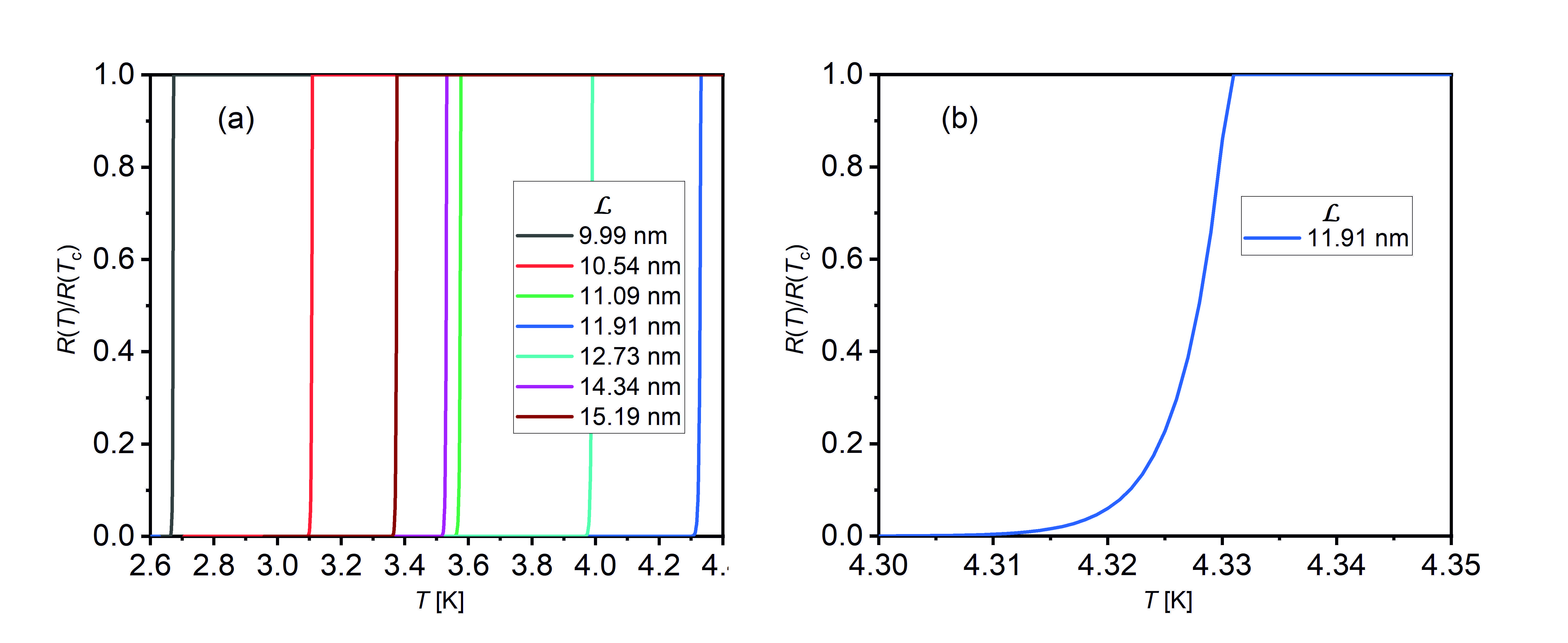}
        \caption{Resistive transitions in quantum rings calculated by Eq. (\ref{Arrhenius}) with critical temperature $T_c$ values taken from the predictions shown in Fig. \ref{critical temperature}a  for $U=0.6$, $D=10$ nm, and the inner radius $a=5$ nm. The parameter $\mathcal{L}$ varies as detailed in the legend in panel (a) and in panel (b). In panel (a) $\mathcal{L}$ grows from left to right as indicated in the legend.}
\label{resistive transitions}
\end{figure*}

The resistive transitions calculated by Eq. (\ref{Arrhenius}) at different values of the parameter $\mathcal{L}$ are represented in Fig. \ref{resistive transitions}a,b. The critical current density $j_c(0)=1.07\times 10^{11}$ Am$^{-2}$ corresponds to Al stripes with the mean free path 15.5 nm \cite{Romijn1982}. The largest resistive transition width $\sim 0.02$ K (see Fig. \ref{resistive transitions}b) occurs at $\mathcal{L}=11.91$ nm, when the critical temperature 4.33 K is maximal.


Analysis of the experimentally accessible quantum interference effects (in particular, Aharonov-Bohm and Aharonov-Casher) peculiar to the doubly-connected topology of quantum rings
\cite{Fomin_book,Fomin_book2}, which requires a generalization of the present approach to include an external magnetic field, will be performed in a forthcoming paper.\\

\section{APPENDIX}

\subsection{Volume of hole pockets and topology of the Fermi sea}
The Fermi energy of the system is related to the volume of the occupied states in momentum space (at $T=0$). We will now proceed to compute this volume as a function of the geometrical parameters of the ring in real space. 

The volume of the allowed or occupied states in k-space can be analytically computed by subtracting the volume of the forbidden states in k-space from the volume of the Fermi sphere:
\begin{equation}
    V_{k} = \frac{4\pi}{3}k_F^3 - V_{holes},
\end{equation}
where $k_F$ is the Fermi wavevector, and $V_{holes}$ is the volume of the forbidden states in k-space.

To this aim, we have to compute $V_{holes}$ and we will divide the problem into multiple sub-cases, as detailed in what follows.

\hfill \break
\begin{center}
    \textbf{1. Fermi radius is greater than both diameters that define the spheres of hole pockets}
\end{center}

In this case (shown in Fig. \ref{fig:proj1}) the volume of the hole pockets is given by the volume of the four spheres minus the volume of the four intersections (denoted as $4V_{intersect}$):
\begin{equation}
    V_{holes} = 2\frac{4\pi}{3}\Big[\Big(\frac{\pi}{\mathcal{L}}\Big)^3+\Big(\frac{\pi}{D}\Big)^3\Big] - 4V_{intersect}.\label{holes1}
\end{equation}
We can determine $V_{intersect}$ exactly by noting that the intersection between two interpenetrating spheres is a circumference, so the intersection is given by two spherical caps with the same base, of radius $k_{cap}$ given by:
\begin{equation}
    k_{cap} = \frac{\pi}{\sqrt{D^2+\mathcal{L}^2}}
\end{equation}
and the corresponding heights are given, respectively by:
\begin{equation}
    \begin{split}
        k_1 & = \frac{2\pi}{\mathcal{L}}\Big(1-\frac{D}{\sqrt{D^2 +\mathcal{L}^2}} \Big) \\
        k_2 & = \frac{2\pi}{D}\Big(1-\frac{\mathcal{L}}{\sqrt{D^2 +\mathcal{L}^2}} \Big). 
    \end{split}
\end{equation}
It then follows that:
\begin{equation}
\begin{split}
   V_{intersect} = \frac{\pi^4}{(b^2-a^2)^{3/2}}\Big\{ \Big[1-\frac{D}{\sqrt{D^2+\mathcal{L}^2}}\Big]^2+\\
   -\frac{1}{3}\Big[1-\frac{D}{\sqrt{D^2+\mathcal{L}^2}}\Big]^3
   \Big\} + \\+  \frac{8\pi^4}{D^3}\Big\{ \Big[1-\frac{\mathcal{L}}{\sqrt{D^2+\mathcal{L}^2}}\Big]^2 + \\ -\frac{1}{3}\Big[1-\frac{\mathcal{L}}{\sqrt{D^2+\mathcal{L}^2}}\Big]^3
   \Big\}.
\end{split}
\end{equation}
By plugging this expression into Eq. \eqref{holes1} we thus obtain the effective volume of the hole pockets in this case.

\begin{center}
    \textbf{2. Fermi radius is greater than the diameter of the intersection spherical cap}
\end{center}
In this case (represented in Figs. \ref{fig:proj2} and \ref{fig:proj3}) the volume of the allowed states is the same as in the previous case except for adding the volume of the parts of the four spheres that lie outside the Fermi sphere (this is necessary because these volumes are subtracted away when subtracting the volume of the four spheres and have to be re-added \emph{ex post facto}). 
The corresponding mathematical condition is therefore
\begin{equation}
    k_{F} < 2k_{sphere_1} \lor k_{F} < 2k_{sphere_2}.
\end{equation}
where $k_{sphere}$ is the radius of the spheres that define the forbidden states (hole pockets).
We also made the additional (reasonable) observation that
\begin{equation}
    2k_{cap} < k_{F}
\end{equation}

When $2k_{cap} > k_{F}$, the volume of allowed states is challenging to compute exactly due to the complex geometry of overlapping spheres and we could not find a solution in this case. We shall consider this model to be applicable to a situation where $2k_{cap} < k_{F}$, which anyway should cover most cases of practical interest. Indeed, the condition $2k_{cap} > k_{F}$ corresponds to situations of extreme confinement (in the order of very few Angstroms) that can hardly be achieved experimentally.

The volume of the four spheres that lies outside the Fermi sphere is given by
 \begin{equation}
 \begin{split}
     V_{outside} = \frac{\pi}{3}\Big\{\Big(2k-\frac{k_{F}^2}{2k}\Big)^2\Big(k+\frac{k_{F}^2}{2k}\Big) + \\ - (k_{F}-\frac{k_{F}^2}{2k}\Big)^2\Big(2k_{F}+\frac{k_{F}^2}{2k}\Big)\Big\}.
 \end{split}
 \end{equation}

For the various cases, the volume of occupied states in k-space can then be calculated as follows:
\begin{itemize}
    \item \text{if} $k_{F} < \frac{2\pi}{\mathcal{L}}$ and $k_{F} > \frac{2\pi}{D}$\\
    \begin{equation}
    V_{k} = \frac{4\pi}{3}k_F^3 - V_{holes} + 2V_{outside}(k=\frac{\pi}{\mathcal{L}} );
    \end{equation}
    
     \item \text{if} $k_{F} > \frac{2\pi}{\mathcal{L}}$ and $k_{F} < \frac{2\pi}{D}$\\
     \begin{equation}
    V_{k} = \frac{4\pi}{3}k_F^3 - V_{holes} + 2V_{outside}(k=\frac{\pi}{D});
    \end{equation}
    \item \text{if} $k_{F} < \frac{2\pi}{\mathcal{L}}$ and $k_{F} < \frac{2\pi}{D}$
    \begin{equation}
\begin{split}
    V_{k} = \frac{4\pi}{3}k_F^3 - V_{holes} + 2V_{outside}(k=\frac{\pi}{D}) +\\+ 2V_{outside}(k=\frac{\pi}{\mathcal{L}} ).
\end{split}
\end{equation}   
\end{itemize}

\subsection*{Acknowledgments} 
A. Z. gratefully acknowledges funding from the European Union through Horizon Europe ERC Grant number: 101043968 ``Multimech'', from US Army Research Office through contract nr. W911NF-22-2-0256, and from the Nieders{\"a}chsische Akademie der Wissenschaften zu G{\"o}ttingen in the frame of the Gauss Professorship program. V. M. F. thanks G. P. Papari and M. V. Fomin for useful discussions.  

\section*{Data availability}
The data that supports the findings of this study are available within the article. 
\section*{Author contribution statement}
E.L. developed calculations under the supervision of A.Z. and V.F. A.Z. and E.L. wrote the paper with input from V.F.

\bibliographystyle{apsrev4-1}
\bibliography{ref}

\end{document}